\documentclass[10pt]{article}
\usepackage{amssymb,amsmath}
\usepackage{amsthm}
\usepackage{mathrsfs}
\usepackage{graphicx}
\usepackage{dcolumn}
\usepackage{bm}
\usepackage{fullpage}
\usepackage{color}
\usepackage[all]{xy}
\usepackage{ulem}
\begin{document}
\newtheorem{Definition}{Definition}[section]
\newtheorem{Theorem}{Theorem}[section]
\newtheorem{Proposition}{Proposition}[section]
\newtheorem{Lemma}{Lemma}[section]
\theoremstyle{definition}
\newtheorem*{Proof}{Proof}
\newtheorem{Example}{Example}[section] 
\newtheorem{Postulate}{Postulate}[section]
\newtheorem{Corollary}{Corollary}[section]
\newtheorem{Remark}{Remark}[section]
\theoremstyle{remark}
\newtheorem*{Claim}{Claim}
\newcommand{\beq}{\begin{equation}}
\newcommand{\beqa}{\begin{eqnarray}}
\newcommand{\eeq}{\end{equation}}
\newcommand{\eeqa}{\end{eqnarray}}
\newcommand{\non}{\nonumber}
\newcommand{\lb}{\label}
\newcommand{\fr}[1]{(\ref{#1})}
\newcommand{\cc}{\mbox{c.c.}}
\newcommand{\nr}{\mbox{n.r.}}
\newcommand{\const}{\mbox{Const.}}
\newcommand{\e}{\mathrm{e}}
\newcommand{\eq}{\mathrm{eq}}
\newcommand{\Id}{\mathrm{I}}
\newcommand{\ddiv}{\mathrm{div}}
\newcommand{\Spec}{\mathrm{Spec}}
\newcommand{\Ising}{\mathrm{Ising}}
\newcommand{\hF}{\widehat F}
\newcommand{\hL}{\widehat L}
\newcommand{\tA}{\widetilde A}
\newcommand{\tB}{\widetilde B}
\newcommand{\tC}{\widetilde C}
\newcommand{\tL}{\widetilde L}
\newcommand{\tK}{\widetilde K}
\newcommand{\tX}{\widetilde X}
\newcommand{\tY}{\widetilde Y}
\newcommand{\tU}{\widetilde U}
\newcommand{\tZ}{\widetilde Z}
\newcommand{\talpha}{\widetilde \alpha}
\newcommand{\te}{\widetilde e}
\newcommand{\tv}{\widetilde v}
\newcommand{\ts}{\widetilde s}
\newcommand{\tx}{\widetilde x}
\newcommand{\ty}{\widetilde y}
\newcommand{\ud}{\underline{\delta}}
\newcommand{\uD}{\underline{\Delta}}
\newcommand{\chN}{\check{N}}
\newcommand{\cA}{{\cal A}}
\newcommand{\cC}{{\cal C}}
\newcommand{\cD}{{\cal D}}
\newcommand{\cF}{{\cal F}}
\newcommand{\cL}{{\cal L}}
\newcommand{\cI}{{\cal I}}
\newcommand{\cM}{{\cal M}}
\newcommand{\cN}{{\cal N}}
\newcommand{\cO}{{\cal O}}
\newcommand{\cQ}{{\cal Q}}
\newcommand{\cS}{{\cal S}}
\newcommand{\cY}{{\cal Y}}
\newcommand{\cU}{{\cal U}}
\newcommand{\cV}{{\cal V}}
\newcommand{\tcA}{\widetilde{\cal A}}
\newcommand{\DD}{{\cal D}}
\newcommand\TYPE[3]{ \underset {(#1)}{\overset{{#3}}{#2}}  }
\newcommand{\bfe}{\boldsymbol e} 
\newcommand{\bfb}{{\boldsymbol b}}
\newcommand{\bfd}{{\boldsymbol d}}
\newcommand{\bfh}{{\boldsymbol h}}
\newcommand{\bfj}{{\boldsymbol j}}
\newcommand{\bfn}{{\boldsymbol n}}
\newcommand{\bfA}{{\boldsymbol A}}
\newcommand{\bfB}{{\boldsymbol B}}
\newcommand{\bfJ}{{\boldsymbol J}}
\newcommand{\dr}{\mathrm{d}}
\newcommand{\sech}{\mathrm{sech}}
\newcommand{\A}{   \TYPE 1  {A}  {}   }
\newcommand{\Ap}{  \TYPE p  {A}  {}   }
\newcommand{\F}{   \TYPE 2  {F}  {}   }
\newcommand{\G}{   \TYPE 2  {G}  {}   }
\newcommand{\hash}{\#}
\newcommand{\hashat}{\widehat{\#}}
\newcommand{\hashch}{\stackrel{\vee}{\#}}
\newcommand{\chd}{\stackrel{\vee}{\D}}
\newcommand{\normM}[2]{\left(  #1\, , \, #2 \right)}
\newcommand{\normU}[2]{\left\{ #1\, , \, #2 \right\}}
\newcommand{\inp}[2]{\left\langle\,  #1\, , \, #2\, \right\rangle}
\newcommand{\rmt}[1]{{\mathrm{t}}({#1})}
\newcommand{\rmo}[1]{{\mathrm{o}}({#1})}
\newcommand{\rmA}{{\mathrm{A}}}
\newcommand{\rmR}{{\mathrm{R}}}
\newcommand{\rmL}{\mathrm{L}}
\newcommand{\wt}[1]{\widetilde{#1}}
\newcommand{\wh}[1]{\widehat{#1}}
\newcommand{\ch}[1]{\check{#1}}
\newcommand{\ol}[1]{\overline{#1}}
\newcommand{\ii}{\imath}
\newcommand{\ic}{\iota}
\newcommand{\mbbE}{\mathbb{E}}
\newcommand{\mbbP}{\mathbb{P}}
\newcommand{\mbbR}{\mathbb{R}}
\newcommand{\mbbN}{\mathbb{N}}
\newcommand{\mbbZ}{\mathbb{Z}}
\newcommand{\Leftrightup}[1]{\overset{\mathrm{#1}}{\Longleftrightarrow}}
\newcommand{\avgg}[2]{\left\langle\,{#1}\, \right\rangle_{#2}}
\newcommand{\spanmath}{\mathrm{span}\,}
\title{Diffusion equations from master equations\\
--- A discrete geometric approach --- }
\author{  Shin-itiro GOTO and Hideitsu HINO\\
The Institute of Statistical Mathematics,\\ 
10-3 Midori-cho, Tachikawa, Tokyo 190-8562, Japan 
}
\date{\today}
\maketitle
\begin{abstract}%
In this paper, continuous-time master equations with finite states 
employed in nonequilibrium statistical mechanics 
are formulated in the language of discrete geometry. 
In this formulation,    
chains in algebraic topology are used, and 
master equations are described on graphs that consist of vertexes 
representing states and of directed edges representing transition matrices. 
It is then shown that master equations under the detailed balance conditions 
are equivalent to discrete diffusion equations, where the 
Laplacians are    
defined as  self-adjoint operators  
with respect to introduced 
inner products.  
An isospectral property  
of these Laplacians is shown for non-zero eigenvalues, 
and its applications are given. The 
convergence to the equilibrium state is shown by analyzing this 
class of diffusion equations.
In addition, a systematic way to derive closed dynamical systems 
for expectation values is given.
For the case that the detailed balance conditions are not imposed,
master equations are expressed as a form of a continuity equation.

\end{abstract}%

%
\section{Introduction}
Master equations are vital in the study of 
nonequilibrium statistical mechanics\,\cite{Kubo1991,Klafter2011,Weber2017}, 
since they are mathematically simple and show relaxation 
processes towards to equilibrium states under some
conditions\,\cite{VanKampen2007}.  
These equations describe 
the time evolution of probabilities,
and 
they are first order differential or difference equations. 
In addition, these equations are used in Monte Carlo 
simulations\,\cite{Binder1997}. 
Quantum mechanical case can be considered by extending 
classical systems\,\cite{Lindblad1976}.
Thus there have been a variety of applications in mathematical sciences, and 
its progress continues to attract attention in the 
literature\,\cite{Baez2018,Goto2019,Goto2019PhysScr,Sakai2013}.

Algebraic topology and graph theory have been applied to 
a variety of sciences and mathematical engineering.
Several topological approaches to master equations exist in the 
literature\,\cite{Schnakenberg1976,Gaspard2007,Ohwa2008,Polettini2012,Polettini2015}.
Not only master equations, but also 
random walks on lattices\,\cite{Sunada2013},  
electric circuits\,\cite{Nakata2016,Zeidler2011,Nakata2019}, and so on, 
have been 
studied from the viewpoint of algebraic topology.
By introducing inner products for functions on graphs
 together with  coboundary operator,   
one can define the adjoint of the coboundary operator,  and 
Laplacians as self-adjoint operators\,\cite{Sunada2013}. 
Corresponding operators in the continuous case 
are useful as proven in the literature of functional 
analysis\,\cite{Yoshida1995}. 
An amalgamation of these mathematical disciplines may be called discrete geometry. 
It is then of interest to explore how above-mentioned operators
 in discrete geometry  
can be used for master equations. 
Moreover,
although discrete diffusion equations have been derived from  
master equations in some cases, the condition when such 
diffusion equations can be derived is not known. 
Since the knowledge of discrete diffusion equations has been accumulated, 
clarifying such a condition is expected to be fruitful 
for the study of master equations.

In this paper continuous-time master equations are formulated in terms of 
functions of chains,  
where   states and transition matrices for master equations  
are described by chains used in algebraic topology.
In particular, probability distribution function is regarded 
as a function of $0$-chain,
and transition matrix a  function of $1$-chain, where 
a discrete state is expressed as a vertex or a $0$-chain.
After introducing some inner products for functions on chains  
and  current as 
a function on $1$-chain, 
master equations are shown to be written in terms of
the adjoint of the coboundary operator acting on 
the current:  
\begin{Claim}
Master equations can be written as a form of a continuity equation 
(See Theorem\,\ref{fact-master-equations-current} for details).
\end{Claim}
\noindent
In addition, under the assumption that the detailed balance conditions hold, 
an equivalence between master equations and discrete diffusion equations 
is shown,  
where the Laplacians are constructed by choosing appropriate measures for 
inner products: 
\begin{Claim}
Master equations under the detailed balance conditions are equivalent to 
discrete diffusion equations 
(See Theorem\,\ref{fact-master-equations-detailed-balance} for details). 
\end{Claim}
\noindent
By applying this statement, 
it is shown that probability distribution functions 
relax to the equilibrium state
(See  
Corollary\,\ref{fact-master-equations-detailed-balance-relaxation} 
and 
Proposition\,\ref{fact-spectrum-decomposition-solution-diffusion}). 
Contrary to this, it is shown that discrete diffusion equations yield 
master equations (See Proposition\,\ref{fact-diffusion-yield-master}).  
Meanwhile, an isospectral property for non-zero eigenvalues of these Laplacians 
is given (See Theorem\,\ref{fact-supersymmetry}), and this can be  
referred to as a supersymmetry\,\cite{Nakahara}. 
With this supersymmetry dynamical systems for 
expectation variables are derived without any approximation
(See Propositions\,\ref{fact-system-from-master-equations-with-a-unique-non-trivial-eigenvalue}--- \ref{fact-dynamical-systems-2}). 
Some inequalities are also derived
(See Propositions\,\ref{fact-inequality-S} and 
\ref{fact-inequality-H-psi}).

These theorems, corollary, and so on 
should be compared with those in the existing literature. 
In the study of random walks on lattices, 
chains, functions and their Laplacians are also used\,\cite{Sunada2013}. 
In the literature 
the probability distribution functions are identified with functions on 
$1$-chains,
which is different to the present formalism. The differences appear 
since transition matrices are introduced for describing state transitions 
for the present study only. In contrast, similarities 
between the present study and existing 
studies appear due to the use of Laplacians. 
Laplacian is a self-adjoint operator with respect to 
an inner product, and brings several properties as well as the case of 
the standard Riemannian geometry.    

The rest of this paper is organized as follows. 
In Section\,\ref{section-Preliminaries}, some preliminaries are 
provided  in order to keep this paper self-contained.
They include boundary operators,
  inner products, and Laplacians. Most of such tools and notions
  have been described 
  in Ref.\,\cite{Sunada2013}, while others are
  invented in this paper. 
In Section\,\ref{section-master-equations}, master equations are formulated 
as dynamical systems on graph, 
and the main claims of this paper  and their consequences are  
provided.
Namely, diffusion equations are derived by means of tools introduced in the previous section. Moreover, various inequalities and dynamical systems for expectation variables are derived.
Section\,\ref{sec-conclusions} 
summarizes this paper and discusses some future studies. 

\section{Preliminaries}
\label{section-Preliminaries}
Let $G=(V,E)$ be a  directed graph with $V$ a vertex set and $E$ an edge set.
Throughout this paper, every graph is finite ($\# E<\infty$), connected, and allowed to have loop edges. 
  In addition, an inverse edge $\ol{e}\in E$ is assumed to exist
  for each edge $e\in E$.   
However parallel edges that will be 
defined later are excluded from 
this contribution.
  Details of these assumptions are explained in
  Section \ref{section-standard-operators}.

\subsection{Standard operators}
\label{section-standard-operators}
In this subsection, most of notions and  notations follow 
Ref.\,\cite{Sunada2013}. However, some variants are also introduced. 

For a given edge $e\in E$, 
the inverse of $e$, the 
terminus of $e$, and the origin of $e$ are denoted by  
$\ol{e}$ , $\rmt{e}\in V$, and $\rmo{e}\in V$, 
respectively, as follows:
$$
\xymatrix@R=2pt{
&\bullet\ar[rr]_{e}&
&\bullet\\
&\rmo{e}&&\rmt{e}
},\qquad\mbox{and}\quad
\xymatrix@R=2pt{
&\bullet&
&\bullet\ar[ll]^{\ol{e}}\\
&\rmt{\ol{e}}&&\rmo{\ol{e}}
},
$$
where $\ol{e}$ has not been depicted in the left graph, and
$e$ has not been depicted in the right graph above.                 
Then, it follows that 
$$
\rmt{\ol{e}}
=\rmo{e},\qquad\mbox{and}\qquad
\rmo{\ol{e}}
=\rmt{e}.
$$
  In this paper $\ol{e}\in E$ is assumed to exist
  for any edge $e\in E$.
  Thus, even $\ol{e}$ is not depicted for a given $e$,  
  the existence of its inverse edge $\ol{e}$ is assumed in this section.
A {\it loop edge} $e\in E$ is such that $\rmo{e}=\rmt{e}$, and 
{\it parallel edges} $e_{\,1},e_{\,2}\in E$ are such that $e_{\,1}\neq e_{\,2}$ with 
$\rmo{e_{\,1}}=\rmo{e_{\,2}}$ and $\rmt{e_{\,1}}=\rmt{e_{\,2}}$. 
Parallel edges are assumed not to exist in any graph in this paper.

Then one defines 
the groups of {\it $0$-chains} and 
{\it $1$-chains} on a graph $G$ with coefficients $\mbbR$ 
\beqa
C_{\,0}(G,\mbbR)
&:=&\left\{\,\sum_{x\in V}a_{\,x}x\ \bigg|\ a_{\,x}\in\mbbR \ \right\},
\non\\
C_{\,1}(G,\mbbR)
&:=&\left\{\,\sum_{e\in E}a_{\,e}e\ \bigg|\ a_{\,e}\in\mbbR \ \right\},
\non
\eeqa
respectively.
Here $\sum_{e\in E}a_{\,e}e$ denotes the sum over all $e\in E$
  in the sense that
  $$
\sum_{e\in E}a_{\,e}e
  =\sum_{\ol{e}\in E}a_{\,\ol{e}}\ol{e}.
  $$
As an example, consider the graph $G=(V,E)$ with 
$$
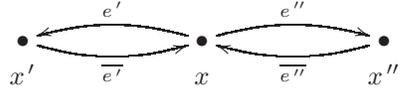
\xymatrix@R=2pt{
&\bullet\ar@/_/[rr]_{\ol{e^{\,\prime}}}
&
&\bullet\ar@/_/[ll]_{e^{\,\prime}}\ar@/^/[rr]^{e^{\,\prime\prime}}
&
&\bullet\ar@/^/[ll]^{\ol{e^{\,\prime\prime}}}\\
&x^{\,\prime}
&
&x
&
&x^{\,\prime\prime}
}.
$$
In this case, with some $a_{\,x^{\,\prime}},a_{\,x},a_{\,x^{\,\prime\prime}}\in\mbbR$, 
an element of $C_{\,0}(G)$ is expressed as 
$$
\sum_{y\in V}a_{\,y}\,y
=a_{\,x^{\,\prime}}\,x^{\,\prime}
+a_{\,x}\,x
+a_{\,x^{\,\prime\prime}}\,x^{\,\prime\prime}.
$$
Then with some 
$a_{\,e^{\,\prime}},a_{\,e^{\,\prime\prime}},a_{\,\ol{e^{\,\prime}}},a_{\,\ol{e^{\,\prime\prime}}}\in\mbbR$, an element of $C_{\,1}(G)$ is expressed as  
$$
\sum_{e\in E}a_{\,e}\,e
=a_{\,e^{\,\prime}}\,e^{\,\prime}
+a_{\,e^{\,\prime\prime}}\,e^{\,\prime\prime}
+a_{\,\ol{e^{\,\prime}}}\,\ol{e^{\,\prime}}
+a_{\,\ol{e^{\,\prime\prime}}}\,\ol{e^{\,\prime\prime}}
=\sum_{\ol{e}\in E}a_{\,\ol{e}}\,\ol{e}.
$$
The spaces of functions on $C^{\,0}(G,\mbbR)$ and 
$C^{\,1}(G,\mbbR)$ are denoted by 
\beqa
C^{\,0}(G,\mbbR)
&:=&\left\{\ f : C_{\,0}(G,\mbbR)\to\mbbR \right\},
\non\\
C^{\,1}(G,\mbbR)
&:=&\left\{\ \omega : C_{\,1}(G,\mbbR)\to\mbbR \right\}.
\non
\eeqa
Elements of the subset of $C^{\,0}(G,\mbbR)$ being dual to $C_{\,0}(G,\mbbR)$ 
are referred to as {\it $0$-cochains}.
Here the dual space of a linear space is, by definition,
 a linear space in general. 
Similarly, 
elements of the subset of $C^{\,1}(G,\mbbR)$ being dual to $C_{\,1}(G,\mbbR)$ 
are referred to as {\it $1$-cochains}.
The set $C^{\,0}(G,\mbbR)$ is also denoted by $\Lambda^{\,0}(G,\mbbR)$ in this 
paper.
  Note that $f(x) \in\mbbR$ with $f\in C^{\,0}(G,\mbbR)$ needs not
  be linear in $x$, and that
  $\omega(e) \in\mbbR$ with $\omega\in C^{\,1}(G,\mbbR)$ needs not be linear in $e$. 
The subsets    
$\Lambda^{\,1}(G,\mbbR)\subset C^{\,1}(G,\mbbR)$ 
and $S^{\,1}(G,\mbbR)\subset C^{\,1}(G,\mbbR)$ are defined by  
\beqa
\Lambda^{\,1}(G,\mbbR)
&:=&\left\{\ \omega\in C^{\,1}(G,\mbbR)\ |\ 
\omega(\,\ol{e}\,) =-\,\omega(e)
 \right\},
\non\\
S^{\,1}(G,\mbbR)
&:=&
\left\{\ \mu\in C^{\,1}(G,\mbbR)\ |\ 
\mu(\,\ol{e}\,) =\mu(e)
 \right\}.
\non
\eeqa
In what follows, $C_{\,0}(G,\mbbR)$ and $C_{\,1}(G,\mbbR)$ 
are often abbreviated as $C_{\,0}(G)$ and $C_{\,1}(G)$, respectively. 
Similar abbreviations will be adopted.

The {\it boundary operator} is defined as 
$$
\partial : C_{\,1}(G)\to C_{\,0}(G),\qquad\mbox{so that}\qquad
\partial (e):=\rmt{e}-\rmo{e},
$$
and the linearity holds: 
$$
\partial (c_{\,1}+c_{\,2})
=\partial c_{\,1}+\partial c_{\,2},\qquad\mbox{and}\qquad
\partial (a\,c)
=a\,\partial\, c,\qquad a\in\mbbR,\quad c,c_{\,1},c_{\,2}\in C_{\,1}(G).
$$
The dual of the boundary operator, the {\it coboundary operator}, is defined by 
$$
\dr :\Lambda^{\,0}(G) \to  
\Lambda^{\,1}(G),\qquad\mbox{so that}\qquad
(\dr f  )(e)
:= f (\rmt{e})- f (\rmo{e}),
$$ 
and the linearity holds: 
$$
\dr (f_{\,1}+f_{\,2})
= \dr f_{\,1}+\dr f_{\,2},\qquad\mbox{and}\qquad
\dr (a\,f)
=a\,\dr f,\qquad a\in\mbbR,\quad f,f_{\,1},f_{\,2}\in \Lambda^{\,0}(G).
$$
If  $f$ is a $0$-cochain, then it follows that 
$(\dr f)(e)=f(\partial e)$.
For any loop edge $e\in E$, it follows from $f(\rmt{e})=f(\rmo{e})$ 
that $(\dr f)(e)=0$.
This $\dr f$ is indeed an element belonging to $\Lambda^{\,1}(G)$, 
as shown below.  
For any $f \in \Lambda^{\,0}(G)$ and $e\in E$, one has that 
$$
(\dr f)\,(\,\ol{e}\,)
=f (\,\rmt{\ol{e}}\,)-f(\,\rmo{\ol{e}}\,)
=f (\,\rmo{e}\,)-f(\,\rmt{e}\,)
=-\,(\dr f)(\,e\,).
$$

To define inner products, 
$$
\inp{}{}_{V}:\Lambda^{\,0}(G)\times \Lambda^{\,0}(G)\to\mbbR
\qquad\mbox{and}\qquad 
\inp{}{}_{E}:\Lambda^{\,1}(G)\times \Lambda^{\,1}(G)\to\mbbR,
$$ 
one introduces some measures.  
Let $m_{\,V}$ and $m_{\,E}$ be 
elements of $C^{\,0}(G)$ and $S^{\,1}(G)$, such that 
$$
m_{\,V}(x)>0,\quad \forall\ x \in V,
$$
and
$$
m_{\,E}(e)
=m_{\,E}(\,\ol{e}\,)>0,\quad\forall\ e \in E.
$$
This $m_{\,E}\in S^{\,1}(G)$ is referred to as  a {\it reversible measure}.

The following are inner products 
\beqa
\inp{f_{\,1}}{f_{\,2}}_{\,V}
&:=&\sum_{x\in V} f_{\,1}(x)\,f_{\,2}(x)\,m_{\,V}(x), \qquad
\forall\ f_{\,1},f_{\,2}\ \in \Lambda^{\,0}(G),
\label{inner-product-vertex}\\
\inp{\omega_{\,1}}{\omega_{\,2}}_{\,E}
&:=&\frac{1}{2}\sum_{e\in E} \omega_{\,1}(e)\,\omega_{\,2}(e)\,m_{\,E}(e),
\qquad \forall\ \omega_{\,1},\omega_{\,2}\ \in \Lambda^{\,1}(G). 
\label{inner-product-edge}
\eeqa
Associated with this set of inner products, one defines the 
{\it adjoint of the coboundary operator}
$\dr^{\dagger}:\Lambda^{\,1}(G)\to\Lambda^{\,0}(G)$. This $\dr^{\dagger}$ is also
referred to as {\it co-derivative} in this paper,
which is explicitly written as  
\beq
(\dr^{\dagger}\omega)(x)
=\frac{-1}{m_{\,V}(x)}\sum_{e\,\in E_{\,x}}\omega(e)\,m_{\,E}(e),
\label{co-derivative}
\eeq 
where 
\beq
E_{\,x}
:=\left\{\,e\in E\ |\ \rmo{e}=x
\,\right\}.
\label{E-x}
\eeq
It follows from \fr{co-derivative} 
that the linearity holds: 
$$
\dr^{\dagger}(\omega_{\,1}+\omega_{\,2})
=\dr^{\dagger}\omega_{\,1}+\dr^{\dagger}\omega_{\,2},\qquad\mbox{and}\qquad
\dr^{\dagger}(a\,\omega )
=a\,\dr^{\dagger}\omega,\qquad 
a\in\mbbR,\quad\omega,\omega_{\,1},\omega_{\,2}\in\Lambda^{\,1}(G). 
$$
Notice that the decomposition of the sum 
\beq
  \sum_{e\in E}\cdots
  =\sum_{x\in V}\sum_{e\in E_{\,x}}\cdots
\label{sum-e-decomposition-vertex-edge-at-x}
  \eeq
  holds. 
To show examples of \fr{E-x} and \fr{sum-e-decomposition-vertex-edge-at-x}, 
consider the graph $G=(V,E)$,  
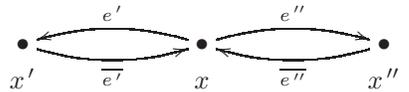
$$
\xymatrix@R=2pt{
&\bullet\ar@/_/[rr]_{\ol{e^{\,\prime}}}
&
&\bullet\ar@/_/[ll]_{e^{\,\prime}}\ar@/^/[rr]^{e^{\,\prime\prime}}
&
&\bullet\ar@/^/[ll]^{\ol{e^{\,\prime\prime}}}\\
&x^{\,\prime}
&
&x
&
&x^{\,\prime\prime}
}.
$$
For all vertexes, \fr{E-x} is expressed as
$$
E_{\,x^{\prime}}=\{\,\ol{e^{\,\prime}}\,\},\qquad
E_{\,x}=\{\,e^{\,\prime}, e^{\,\prime\prime}\,\},\qquad
E_{\,x^{\prime\prime}}=\{\,\ol{e^{\,\prime\prime}}\,\},
$$
and the sum \fr{sum-e-decomposition-vertex-edge-at-x}
with $\omega\in\Lambda^{\,1}(G)$ 
is expressed as
\beqa
\sum_{e\in E}\omega(e)
&=&
\omega(\,\ol{e^{\,\prime}}\,)
+\left[\,\omega(\,e^{\,\prime}\,)+\omega(\,e^{\,\prime\prime}\,)\,\right]
+\omega(\,\ol{e^{\,\prime\prime}}\,)
\non\\
&=&
\sum_{e\in E_{\,x^{\,\prime}}}\omega(e)
+\sum_{e\in E_{\,x}}\omega(e)
+\sum_{e\in E_{\,x^{\prime\prime}}}\omega(e)
=\sum_{y\in V}\sum_{e\in E_{\,y}}\omega(e).
\non
\eeqa
The operator  $\dr^{\,\dagger}$ is indeed  
the adjoint one of $\dr$ as shown below. 
\begin{Lemma}
\label{Fact-d-d-dagger-1}
$$
\inp{\dr f}{\omega}_{\,E}
=\inp{f}{\dr^{\dagger}\omega}_{\,V}.
$$
\end{Lemma}
\begin{Proof}
  Straightforward calculations with
    \fr{sum-e-decomposition-vertex-edge-at-x}  yield 
\beqa
\inp{\dr f}{\omega}_{\,E}
&=&\frac{1}{2}\sum_{e\in E}(\dr f)(e)\omega(e)\,m_{\,E}(e)
=\frac{1}{2}\sum_{e\in E}\left[\,
f(\rmt{e})-f(\rmo{e})\,\right]\,\omega(e)\,m_{\,E}(e)
\non\\
&=&\frac{-1}{2}\sum_{e\in E}\left[\,
f(\rmo{e})\,\omega(e)-f(\rmt{e})\,\omega(e)\,\right]\,m_{\,E}(e)
\non\\
&=&\frac{-1}{2}\sum_{e\in E}\left[\,
f(\rmo{e})\,\omega(e)\,m_{\,E}(e)
+f(\rmo{\ol{e}})\,\omega(\,\ol{e}\,)\,m_{\,E}(\,\ol{e}\,)\,\right]
\non\\
&=&
      -\,\sum_{e\in E}f(\rmo{e})\,\omega(e)\,m_{\,E}(e)
\non\\
&=&
\sum_{x\in V}\sum_{e\in E_{\,x}}f(x)\,\left[\frac{-1}{m_{\,V}(x)}\omega(e)\,m_{\,E}(e)\right]\,m_{\,V}(x)
\non\\
&=&\sum_{x\in V}f(x)\,(\,\dr^{\dagger}\omega\,)(x)\,m_{\,V}(x)
=\inp{f}{\dr^{\dagger}\omega}_{\,V}.
\non
\eeqa
\qed
\end{Proof}
\begin{Remark}
\label{remark-divergence}
An 
operator analogous to 
$-\,\dr^{\dagger}$ 
in the continuous 
standard Riemannian geometry 
is referred to as divergence. 
Thus, $-\,\dr^{\dagger}$ can be referred to as 
{\it divergence } on graph\,\cite{Jiang2011}.
This operator $\Lambda^{\,1}(G)\to \Lambda^{\,0}(G)$ is denoted by 
\beq
\ddiv
:=-\,\dr^{\dagger}.
\label{definition-divergence}
\eeq
\end{Remark}

The {\it Laplacian acting on $\Lambda^{\,0}(G)$}, 
$\Delta_{\,V}:\Lambda^{\,0}(G)\to\Lambda^{\,0}(G)$,  
is defined as 
\beq  
\Delta_{\,V}
:=-\,\dr^{\dagger}\dr.
\label{Delta0}
\eeq  
The explicit form of its action is obtained as follows. By 
putting $\omega=\dr f$ for $f\in\Lambda^{\,0}(G)$, one has
\beqa
(\Delta_{\,V} f)(x)
&=&-\,(\dr^{\dagger}\dr f)(x)
=-\,(\dr^{\dagger}\,\omega)(x)
\non\\
&=&\frac{1}{m_{\,V}(x)}\sum_{e\,\in E_{\,x}}\omega(e)\,m_{\,E}(e)
=\frac{1}{m_{\,V}(x)}\sum_{e\,\in E_{\,x}}(\,\dr f\,)(e)\,m_{\,E}(e).
\non
\eeqa
Since $\dr$ and $\dr^{\dagger}$ are linear operators, 
$\Delta_{\,V}$ is  
also a linear operator: 
$$
\Delta_{\,V}(f_{\,1}+f_{\,2})
=\Delta_{\,V}f_{\,1}+ \Delta_{\,V} f_{\,2},\qquad\mbox{and}\qquad
\Delta_{\,V}(a\,f )
=a\,\Delta_{\,V}f,\qquad a\in\mbbR,\ f,f_1,f_2\in\Lambda^{\,0}(G).  
$$
Moreover, it follows that 
\beq
(\Delta_{\,V} f)(x)
=\frac{1}{m_{\,V}(x)}\sum_{e\,\in E_{\,x}}
\left[\,f(\,\rmt{e}\,)-f(\,\rmo{e}\,)\,\right]\,m_{\,E}(e).
\label{Delta0-explicit-form}
\eeq

To see a link between \fr{Delta0-explicit-form} and 
a well-known form of discrete Laplacian,
choose $m_{\,V}(x)=\delta$ and $m_{\,E}(e)=1$ for all $x\in V$ and $e\in E$,
where $\delta>0$ is a constant. Then consider the graph 
$$
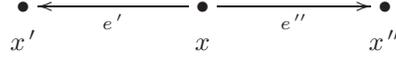
\xymatrix@R=2pt{
&\bullet
&
&\bullet\ar[ll]^{e^{\,\prime}}\ar[rr]_{e^{\,\prime\prime}}
&
&\bullet\\
&x^{\,\prime}
&
&x
&
&x^{\,\prime\prime}
}
$$
where $\ol{e^{\,\prime}}$ and $\ol{e^{\,\prime\prime}}$
have been omitted.  
For this graph, one has the well-known form: 
$$
(\Delta_{\,V} f)(x)
=\frac{1}{\delta}\left[\,
f(x^{\,\prime})-2 f(x)+ f(x^{\,\prime\prime})
\,\right].
$$
By taking the limit $\delta\to 0$ appropriately, 
one has the second derivative of $f$. See Ref.\,\cite{Sunada2008}
for another link between this Laplacian $\Delta_{\,V}$ and    
the so-called adjacency operator.

The operator $\Delta_{\,V}$ is self-adjoint as shown below.
\begin{Lemma}
\label{fact-Delta0-self-adjoint}
$$
\inp{\Delta_{\,V} f_{\,1}}{f_{\,2}}_{\,V}
=\inp{f_{\,1}}{\Delta_{\,V} f_{\,2}}_{\,V}.
$$
\end{Lemma}
\begin{Proof}
It follows that 
$$
\inp{\Delta_{\,V} f_{\,1}}{f_{\,2}}_{\,V}
=\inp{-\,\dr^{\dagger}\,\dr f_{\,1}}{f_{\,2}}_{\,V}
=\inp{\dr f_{\,1}}{-\,\dr\,f_{\,2}}_{\,E}
=\inp{f_{\,1}}{-\,\dr^{\dagger}\, \dr\,f_{\,2}}_{\,V}
=\inp{f_{\,1}}{\Delta_{\,V} f_{\,2}}_{\,V}.
$$
\qed
\end{Proof}

The {\it Laplacian acting on $\Lambda^{\,1}(G)$}, 
$\Delta_{\,E}:\Lambda^{\,1}(G)\to\Lambda^{\,1}(G)$,  
is defined as 
\beq
\Delta_{\,E}
:=-\,\dr\,\dr^{\dagger}.
\label{Delta1}
\eeq 
This operator is self-adjoint as shown below.
\begin{Lemma}
$$
\inp{\Delta_{\,E}\, \omega_{\,1}}{\omega_{\,2}}_{\,E}
=\inp{\omega_{\,1}}{\Delta_{\,E}\, \omega_{\,2}}_{\,E}.
$$
\end{Lemma}
\begin{Proof}
It can be proven by straightforward calculations.
To this end, we put $f_{\,1}=\dr^{\,\dagger}\omega_1\in\Lambda^{\,0}(G)$ and 
then it follows that 
\beqa
\inp{\Delta_{\,E}\, \omega_{\,1}}{\omega_{\,2}}_{\,E}
&=&\inp{-\,\dr\,\dr^{\dagger}\, \omega_{\,1}}{\omega_{\,2}}_{\,E}
=\inp{-\,\dr f_{\,1}}{\omega_{\,2}}_{\,E}
=\inp{ f_{\,1}}{-\,\dr^{\dagger}\,\omega_{\,2}}_{\,V}
\non\\
&=&
\inp{\dr^{\dagger}\omega_{\,1}}{-\,\dr^{\dagger}\, \omega_{\,2}}_{\,V} 
=\inp{\omega_{\,1}}{-\,\dr\,\dr^{\dagger}\, \omega_{\,2}}_{\,E}
=\inp{\omega_{\,1}}{\Delta_{\,E}\, \omega_{\,2}}_{\,E}.
\non
\eeqa
\qed
\end{Proof}

Most of the operators and their properties discussed so far 
are well-known\,\cite{Sunada2013}.
By contrast, those discussed in the 
next subsection are not standard ones.
\subsection{Operators for master equations}
To discuss how to describe  master equations 
in the language of discrete geometry, one first 
introduces  
\beq
C_{\,\rmR}^{\,1}(G,\mbbR)
:=\left\{\ \varphi\in C^{\,1}(G,\mbbR)\ \bigg|\ 
\varphi(\,\ol{e}\,)
=\frac{1}{\varphi(e)}\right\}.
\label{condition-for-varphi}
\eeq
Then, with a prescribed $\varphi\in C_{\,\rmR}^{\,1}(G)$, one introduces  
$$
C_{\,\varphi}^{\,1}(G,\mbbR)
:=\left\{\ \omega\in C^{\,1}(G,\mbbR)\ |\ 
\omega(\ol{e})
=-(\varphi(e))^{\,-1}\,\omega(e),
\quad\varphi\in C_{\,\rmR}^{\,1}(G,\mbbR)
\ \right\}.
$$
The operator associated with $\varphi\in C_{\,\rmR}^{\,1}(G)$ is defined as  
\beq
\dr_{\,\varphi}: 
C^{\,0}(G,\mbbR)\to C_{\,\varphi}^{\,1}(G,\mbbR),
\qquad
\mbox{such that}
\qquad
(\dr_{\,\varphi}\,f)(e)
:=\varphi(e)\,f(\rmt{e})-f(\rmo{e}),
\label{d-varphi}
\eeq
and the linearity in $f$ holds:
$$
\dr_{\,\varphi}\,(f_{\,1}+f_{\,2})
=\dr_{\,\varphi}\,f_{\,1}+\dr_{\,\varphi}\,f_{\,2},\qquad\mbox{and}\qquad
\dr_{\,\varphi}\,(a f)
=a\,\dr_{\,\varphi}\, f, \qquad a\in\,\mbbR,\quad f,f_{\,1},f_{\,2}\in\,C^{\,0}(G).
$$ 
Condition in \fr{condition-for-varphi} 
is to guarantee the property 
$\dr_{\,\varphi}\,f\in C_{\,\varphi}^{\,1}(G)$ for 
$f\in C^{\,0}(G)$,
$$
(\dr_{\,\varphi}\,f)(\,\ol{e}\,)
=\frac{-1}{\varphi(e)}(\dr_{\,\varphi}\,f)(\,e\,).
$$
The above equality is verified as 
\beqa
(\dr_{\,\varphi}\,f)(\ol{e})
&=&\varphi(\ol{e})\,f(\rmt{\ol{e}})-f(\rmo{\ol{e}})
=\varphi(\ol{e})\,f(\rmo{e})-f(\rmt{e})
=\frac{1}{\varphi(e)}
f(\rmo{e})-f(\rmt{e})
\non\\
&=&\frac{-1}{\varphi(e)}(\dr_{\,\varphi}\,f\,)(e).
\non
\eeqa
\begin{Remark}
\label{fact-d-phi-becomes-the-standard-d}
In the case that $\varphi$ is such that $\varphi(e)=1$ for any $e\in E$, 
one has that 
$\dr_{\,\varphi}\,f=\dr f$ for any $f\in C^{\,0}(G)$.
The operator $\dr_{\,\varphi}$ is an analogue of the one introduced in 
Ref.\,\cite{Higuchi2001}.
\end{Remark}

As well as the case for $\inp{}{}_{\,E}$, one introduces an 
{\it inner product on $C_{\,\varphi}^{\,1}(G)$}
$$
\inp{}{}_{\,E}^{\varphi}: C_{\,\varphi}^{\,1}(G)\times
C_{\,\varphi}^{\,1}(G)\to \mbbR,
$$  
associated with a reversible measure $m_{\,E}\in S^{\,1}(G)$.
The value of the inner product is the same as \fr{inner-product-edge}: 
\beq
  \inp{\omega_{\,1}}{\omega_{\,2}}_{\,E}^{\varphi}
  :=\frac{1}{2}\sum_{e\in E}\omega_{\,1}(e)\,\omega_{\,2}(e)\,m_{\,E}(e).
\label{inner-product-edge-varphi}
\eeq
The inner product \fr{inner-product-edge-varphi}
      should be notationally distinguished from
  \fr{inner-product-edge}, since \fr{inner-product-edge}
  has been defined as $\inp{}{}_{\,E}:\Lambda^{\,1}(G)\times \Lambda^{\,1}(G)\to \mbbR$. 
The inner product $\inp{}{}_{\,E}^{\varphi}$ coincides with $\inp{}{}_{\,E}$
  when $\varphi(e)=1$ for all $e\in E$.
    With \fr{inner-product-edge-varphi}, one introduces the 
{\it co-derivative on $C_{\,\rmR}^{\,1}(G)$}, 
$\dr_{\,\varphi}^{\dagger}: C_{\,\varphi}^{\,1}(G)\to C^{\,0}(G)$ with 
$\varphi\in C_{\,\rmR}^{\,1}(G)$ 
such that 
$$
(\dr_{\,\varphi}^{\dagger}\,\omega)(x)
:=\frac{-1}{m_{\,V}(x)}\sum_{e\in E_{\,x}}\omega(e)\,m_{\,E}(e).
$$
\begin{Lemma}
For $\varphi\in C_{\,\rmR}^{\,1}(G),\ \omega\in C_{\,\varphi}^{\,1}(G),$ and 
$f\in C^{\,0}(G)$, one has
$$
\inp{\dr_{\,\varphi}f}{\omega}_{\,E}^{\varphi}
=\inp{f}{\dr_{\,\varphi}^{\dagger}\omega}_{\,V},\qquad
$$
\end{Lemma}
\begin{Proof}
Substituting
$$
\varphi(e)
=\frac{1}{\varphi(\,\ol{e}\,)},\qquad
f(\rmt{e})=f(\rmo{\ol{e}}),\qquad
\omega(e)
=-\,\varphi(\,\ol{e}\,)\,\omega(\,\ol{e}\,),\qquad\mbox{and}\qquad
m_{\,E}(e)
=m_{\,E}(\,\ol{e}\,),
$$ 
into 
\beqa
\inp{\dr_{\,\varphi}f}{\omega}_{\,E}^{\varphi}
&=&\frac{1}{2}\sum_{e\in E}\left[\,
\varphi(e)f(\rmt{e})-f(\rmo{e})
\,\right]\,\omega(e)\,m_{\,E}(e)
\non\\
&=&
\frac{-\,1}{2}\sum_{e\in E}\left[\,
f(\rmo{e})\,\omega(e)-\varphi(e)f(\rmt{e})\,\omega(e)
\,\right]\,m_{\,E}(e),
\non
\eeqa
one has
\beqa
\inp{\dr_{\,\varphi}f}{\omega}_{\,E}^{\varphi}
&=&\frac{-\,1}{2}\sum_{e\in E}\left[\,
f(\rmo{e})\,\omega(e)-\frac{1}{\varphi(\,\ol{e}\,)}f(\rmo{\ol{e}})\,
(-\varphi(\,\ol{e}\,)\,\omega(\,\ol{e}\,)\,)
\,\right]\,m_{\,E}(e)
\non\\
&=&
\frac{-\,1}{2}\sum_{e\in E}\left[\,
f(\rmo{e})\,\omega(e)\,m_{\,E}(e)+f(\rmo{\ol{e}})\,
\omega(\,\ol{e}\,)\,m_{\,E}(\ol{e})
\,\right]
=\inp{f}{\dr_{\,\varphi}^{\dagger}\,\omega}_{\,V}.
\non
\eeqa
\qed
\end{Proof}

Although the following operators will not be used in 
Section \,\ref{section-master-equations}, the
Laplacians are  
discussed for the sake of completeness. 
The Laplacian $\Delta_{\,V}^{\,\varphi} : C^{\,0}(G)\to C^{\,0}(G)$ 
is defined as 
$$
\Delta_{\,V}^{\,\varphi}
:=-\,\dr_{\,\varphi}^{\dagger}\,\dr_{\,\varphi}.
$$ 
The explicit form of its action is obtained as follows. By 
putting $\omega=\dr_{\,\varphi} f$ for $f\in C^{\,0}(G)$, one has
\beqa
(\Delta_{\,V}^{\,\varphi} f)(x)
&=&-\,(\dr_{\,\varphi}^{\dagger}\,\dr_{\,\varphi}\, f)(x)
=-\,(\dr_{\,\varphi}^{\dagger}\,\omega)(x)
\non\\
&=&\frac{1}{m_{\,V}(x)}\sum_{e\,\in E_{\,x}}\omega(e)\,m_{\,E}(e)
=\frac{1}{m_{\,V}(x)}\sum_{e\,\in E_{\,x}}(\,\dr_{\,\varphi}\, f\,)(e)\,m_{\,E}(e)
\non\\
&=&
\frac{1}{m_{\,V}(x)}\sum_{e\,\in E_{\,x}}
\left[\,\varphi(e)\,f(\,\rmt{e}\,)-f(\,\rmo{e}\,)\,\right]\,m_{\,E}(e).
\non
\eeqa
This operator is self-adjoint:
$$
\inp{\Delta_{\,V}^{\,\varphi} f_{\,1}}{f_{\,2}}_{\,V}
=\inp{f_{\,1}}{\Delta_{\,V}^{\,\varphi} f_{\,2}}_{\,V},
$$
since 
$$
\inp{\Delta_{\,V}^{\,\varphi} f_{\,1}}{f_{\,2}}_{\,V}
=\inp{-\,\dr_{\,\varphi}^{\dagger}\,\dr_{\,\varphi}\, f_{\,1}}{f_{\,2}}_{\,V}
=\inp{\dr_{\,\varphi} f_{\,1}}{-\,\dr_{\,\varphi}\,f_{\,2}}_{\,E}^{\,\varphi}
=\inp{f_{\,1}}{-\,\dr_{\,\varphi}^{\dagger}\, \dr_{\,\varphi}\,f_{\,2}}_{\,V}
=\inp{f_{\,1}}{\Delta_{\,V}^{\varphi} f_{\,2}}_{\,V}.
$$

In addition, the Laplacian
$\Delta_{\,E}^{\,\varphi} : C_{\,\varphi}^{\,1}\to C_{\,\varphi}^{\,1}$ 
is defined as
$$
\Delta_{\,E}^{\,\varphi}
:=-\,\dr_{\,\varphi}\,\dr_{\,\varphi}^{\dagger}.
$$ 
This operator is self-adjoint:
$$
\inp{\Delta_{\,E}^{\,\varphi}\, \omega_{\,1}}{\omega_{\,2}}_{\,E}^{\,\varphi}
=\inp{\omega_{\,1}}{\Delta_{\,E}^{\,\varphi}\, \omega_{\,2}}_{\,E}^{\,\varphi},
$$
since 
$$
\inp{\Delta_{\,E}^{\,\varphi}\, \omega_{\,1}}{\omega_{\,2}}_{\,E}^{\,\varphi}
=\inp{-\,\dr_{\,\varphi}\,\dr_{\,\varphi}^{\dagger}\, \omega_{\,1}}{\omega_{\,2}}_{\,E}^{\,\varphi}
=\inp{\dr_{\,\varphi}^{\dagger}\, \omega_{\,1}}{-\,\dr_{\,\varphi}^{\dagger}\,\omega_{\,2}}_{\,V}
=\inp{\omega_{\,1}}{-\,\dr_{\,\varphi}^{\dagger}\, \dr_{\,\varphi}\,\omega_{\,2}}_{\,E}^{\,\varphi}
=\inp{\omega_{\,1}}{\Delta_{\,E}^{\varphi}\,\omega_{\,2}}_{\,E}^{\,\varphi}.
$$
\section{Master equations}
\label{section-master-equations}
Master equations  
are used to model nonequilibrium time-development in various 
systems\,\cite{VanKampen2007,Schnakenberg1976}, and they 
are written as 
\beq
\frac{\dr}{\dr t}p_{\,t}(x)
=-\sum_{x'(\neq x)}w_{\,x\to x'}\,p_{\,t}(x)
+\sum_{x'(\neq x)}w_{\,x'\to x}\,p_{\,t}(x'),
\label{standard-master-equations}
\eeq
where $\{x\}$ are discrete states,
$p_{\,t}(x)$ a probability distribution function of 
 $x$ at time $t\in \mbbR$,  
$w_{\,x\to x^{\prime}}$ a transition matrix that 
describes the transition rate from a state $x$ to another state 
$x^{\,\prime}$. In addition, the set
$\{w_{x\to x^{\prime}}\}$ is chosen so 
  that the conservation law $\sum_{x}p_{\,t}(x)=1$ holds, and is used to model
  physical process of a system under consideration. In this paper
we focus on the case where $\{w_{x\to x^{\prime}}\}$ does not depend on time. 
The equilibrium state at $x$ is denoted by 
$p^{\,\eq}(x)$. In what follows, these objects, $\{x\}$, $\{w_{x\to x^{\prime}}\}$,  
and $\{p^{\,\eq}(x)\}$, are treated as given data.
Although there are several closely related equations of 
\fr{standard-master-equations}, including
backward master equations, discrete-time master equations,
and so on\,\cite{Weber2017},  
they are not addressed in this paper.
  
In this section,  
a graph formulation of master equations is shown in terms of objects developed 
in Section\,\ref{section-Preliminaries}. 
Main claims of this paper and their consequences are then provided.

\subsection{Graph formulation}
Introduce a graph $G=(V,E)$ and $w\in C^{\,1}(G)$
associated with the given data 
$\{x\}$ and $\{w_{\,x\to x^{\prime}}\}$ such that the following hold : 
\begin{itemize}
\item
For any $x$, $x$ is an element of $V$, that is, $x\in V$. 
\item
Let $w\in C^{\,1}(G)$ be such that   
$w(e)=w_{\,x\to x^{\prime}}$ for $e\in E$ with 
$\rmt{e}=x^{\,\prime}$ and $\rmo{e}=x$.
\item
For any $e\in E$, its inverse exists, that is, $\ol{e}\in E$. 
If $w(\ol{e})=w_{\,y\to x}$ does not exist, 
then let $w$ be such that  $w(\ol{e})=0$. 
\end{itemize}
  
Then, regard $p_{\,t}$  as follows :
\begin{itemize}
\item
$p_{\,t}\in \Lambda^{\,0}(G)$ so that 
$p_{\,t}(x)\in\mbbR$ for any $x\in V$. 
\end{itemize}

The master equations \fr{standard-master-equations}
can be written in terms of these graph terminologies as 
\beq
\frac{\dr}{\dr t}p_{\,t}(x)
=-\,\sum_{e\in E_{\,x}}
\,p_{\,t}(\rmo{e})\,w(e)+
\sum_{e\in E^{\,x}}
p_{\,t}(\rmo{e})\,w(\,e\,),
\label{homology-master-equations-general}
\eeq
where 
$$
E^{\,x}
:=\{\ e\in E\ |\ \rmt{e}=x\ \}.
$$
The correspondence between \fr{homology-master-equations-general} and
  \fr{standard-master-equations} is obvious. Vertexes $\rmo{e}$ and $\rmt{e}$
  in \fr{homology-master-equations-general} correspond to states $x$
  and $x^{\,\prime}$ in \fr{standard-master-equations},
  and the first    
  $w(e)\in\mbbR$ and second $w(e)\in\mbbR$
  in \fr{homology-master-equations-general} correspond to
  $w_{\,x\to x^{\,\prime}}$ and $w_{\,x^{\,\prime}\to x}$ in
  \fr{standard-master-equations}, respectively.
     Notice that the conservation law
    for \fr{standard-master-equations} translates into  
    $\sum_{x\in V}p_{\,t}(x)=1$ for \fr{homology-master-equations-general}.
  By construction of the present graph formulation, 
  \fr{homology-master-equations-general} reduces further.
  Given an $x\in V$ and $e\in E^{\,x}$,
  there exists $e^{\,\prime}\in E_{\,x}$, such that 
$$
x=\rmt{e}
=\rmo{e^{\,\prime}},
\qquad\mbox{i.e.,}
\xymatrix@R=2pt{
&\bullet\ar@/^/[r]^{e}&\bullet\ar@/^/[l]^{e^{\,\prime}}\\
&x^{\,\prime}&x
}.
$$
It then follows that $\rmo{e}=\rmt{e^{\,\prime}}$ and $e=\ol{e^{\,\prime}}$,  
from which 
$$
\sum_{e\in E^{\,x}}p_{\,t}(\rmo{e})\,w(e)
=\sum_{e^{\,\prime}\in E_{\,x}}p_{\,t}(\rmt{e^{\prime}})\,w(\,\ol{e^{\,\prime}}\,).
$$
Then \fr{homology-master-equations-general} reduces to  
\beq
\frac{\dr}{\dr t}p_{\,t}(x)
=\sum_{e\in E_{\,x}}
\left[\,
-\,p_{\,t}(\rmo{e})\,w(e)+p_{\,t}
(\rmt{e})\,w(\,\ol{e}\,)
\,\right].
\label{homology-master-equations}
\eeq
  
An example of this set of the procedures is given below.
\begin{Example}
(A two-state model). 
  Consider the set of master equations whose total number of states is $2$, 
$$
\frac{\dr}{\dr t}p_{\,t}(x_{\,1})
=-w_{\,1\to 2}\,p_{\,t}(x_{\,1}),\qquad
\frac{\dr}{\dr t}p_{\,t}(x_{\,2})
=w_{\,1\to 2}\,p_{\,t}(x_{\,1}),
$$
where $w_{\,1\to 2}$ is a positive constant.  
This set of equations can be represented 
in terms of the graph
$$
\xymatrix@R=2pt{
&\bullet\ar@/^/[r]^{e^{\,\prime}}&\bullet\ar@/^/[l]^{e^{\,\prime\prime}}\\
&x_{\,1}&x_{\,2}
},\qquad \mbox{where}\quad
w(\,e^{\,\prime}\,)
=w_{\,1\to 2},\quad\mbox{and}\quad
w(\,e^{\,\prime\prime}\,)
=0,
$$
where the edge $e^{\,\prime\prime}$ has been added such that
  $\ol{e^{\,\prime}}$ exists. 
A consistency between this graph expression and the given 
set of master equations is verified as follows.  
Since $\{e^{\,\prime}\}=E_{\,x_1}$,$\{e^{\,\prime\prime}\}=E_{\,x_2}$, 
$x_{\,1}=\rmo{e^{\,\prime}}=\rmt{e^{\,\prime\prime}}$,  
$x_{\,2}=\rmt{e^{\,\prime}}=\rmo{e^{\,\prime\prime}}$, and 
$\ol{e^{\,\prime\prime}}=e^{\,\prime}$,  
  \fr{homology-master-equations} yields 
$$
\frac{\dr}{\dr t}p_{\,t}(x_{\,1})
=\left[\,
-\,p_{\,t}(\rmo{e^{\,\prime}})\,w(e^{\,\prime})
+p_{\,t}(\rmt{e^{\,\prime}})\,w(\,\ol{e^{\,\prime}}\,)
\,\right]
=-p_{\,t}(x_{\,1})w_{\,1\to 2},
$$
and
$$
\frac{\dr}{\dr t}p_{\,t}(x_{\,2})
=\left[\,
-\,p_{\,t}(\rmo{e^{\,\prime\prime}})\,w(e^{\,\prime\prime})
+p_{\,t}(\rmt{e^{\,\prime\prime}})\,w(\,\ol{e^{\,\prime\prime}}\,)
\,\right]
=p_{\,t}(x_{\,1})w_{\,1\to 2}.
$$
  For completeness, the solution satisfying the conservation law
  $p_{\,t}(x_{\,1})+p_{\,t}(x_{\,2})=1$ for all $t\in\mbbR$ is explicitly
 shown as
\beqa
p_{\,t}(x_{\,1})
&=&
p_{\,0}(x_{\,1})\,\e^{\,-\,w_{\,1\to 2}\,t}
\non\\
p_{\,t}(x_{\,2})
&=&
p_{\,0}(x_{\,1})+p_{\,0}(x_{\,2})-p_{\,0}(x_{\,1})\,\e^{\,-\,w_{\,1\to 2}\,t},
\non
\eeqa
where $p_{\,0}(x_{\,1})$ and $p_{\,0}(x_{\,2})$ are
constants satisfying $0\leq p_{\,0}(x_{\,1})\leq 1$ and
$p_{\,0}(x_{\,2})=1-p_{\,0}(x_{\,1})$.
\end{Example}
In what follows, master equations  
are rewritten
in the language discussed in Section\,\ref{section-Preliminaries}.  
To this end, one defines $J_{\,t}\in C^{\,1}(G)$ and 
$\cI_{\,t}\in \Lambda^{\,1}(G)$ so that 
\beqa
J_{\,t}(e)
&:=&-\,p_{\,t}(\,\rmo{e}\,)\,w(e),
\non\\
\cI_{\,t}(e)
&:=&-J_{\,t}(e)+J_{\,t}(\,\ol{e}\,).
\non
\eeqa
One verifies that $\cI_{\,t}(\ol{e})=-\cI_{\,t}(e)$ immediately,
and that $-\cI_{\,t}(e)$ is the summand in the right hand side of 
\fr{homology-master-equations}, 
$$
-\cI_{\,t}(e)
=J_{\,t}(e)-J_{\,t}(\,\ol{e}\,)
=-\,p_{\,t}(\,\rmo{e}\,)\,w(e)
+\,p_{\,t}(\,\rmo{\ol{e}}\,)\,w(\,\ol{e}\,)
=-\,p_{\,t}(\,\rmo{e}\,)\,w(e)
+\,p_{\,t}(\,\rmt{e}\,)\,w(\,\ol{e}\,),
$$
and thus,
$$
\frac{\dr}{\dr t}p_{\,t}(x)
=-\sum_{e\in E_{\,x}}\cI_{\,t}(e).
$$
The physical meaning of the case of $\cI_{\,t}(e)=0$ 
for a given $e\in E$ is that 
the probability flow or current is locally balanced.  
If the condition $w(\ol{e})=w(e)$ is satisfied for all $e\in E$, then  
one has
\beq
\cI_{\,t}(e)
=-\,w(e)\,(\dr p_{\,t})(e).
\label{Fick-law-general}
\eeq
This relation corresponds to Fick's law of diffusion 
  in continuous media\,\cite{Kubo1991}.  

To describe some measures, introduce 
$$
1_{\,V}\in\Lambda^{\,0}(G),\ \mbox{ and }\ 
1_{\,E}\in S^{\,1}(G)\qquad\mbox{ such that}\qquad
1_{\,V}(x)=1\ \mbox{ and}\ 1_{\,E}(e)=1,\quad 
\forall\, x\in V,\ \forall e\in E.
$$ 
In terms of these objects, one has the following:
\begin{Theorem}
(Continuity equation as  master equations).  
\label{fact-master-equations-current}
Choose $m_{\,V}\in \Lambda^{\,0}(G)$ and $m_{\,E}\in S^{\,1}(G)$ 
to be 
\beq
m_{\,V}
=1_{\,V},\qquad\mbox{and}\qquad
m_{\,E}
=1_{\,E}.
\label{trivial-measures}
\eeq
Then 
\fr{homology-master-equations}
is identical to 
\beq
\frac{\dr}{\dr t}p_{\,t}
=\dr^{\dagger}\cI_{\,t}.
\label{cohomology-master-equations}
\eeq
\end{Theorem}
\begin{Proof}
With the choices $m_{\,V}$ and $m_{\,E}$, 
the co-derivative $\dr^{\dagger}$ defined in
\fr{co-derivative} reduces to the one such that 
$$
(\,\dr^{\dagger}\omega\,)(x)
=-\sum_{e\in E_{\,x}} \omega(e),\qquad \omega\in \Lambda^{\,1}(G).
$$  
From this and \fr{homology-master-equations}, one has
$$
\frac{\dr}{\dr t}p_{\,t}(x)
=-\sum_{e\in E_{\,x}} \cI_{\,t}(e)
=(\dr^{\dagger}\cI_{\,t})(x),
$$
for any $x\in V$. Thus, the desired expression is obtained. 
\qed
\end{Proof}
It is worth mentioning that \fr{cohomology-master-equations} is written 
as a form of a continuity equation in continuum mechanics
\beq
\frac{\dr}{\dr t}p_{\,t}
+\ddiv\, \cI_{\,t}
=0, 
\label{cohomology-master-equations-divergence}
\eeq
where $\ddiv:\Lambda^{\,1}(G)\to \Lambda^{\,0}(G)$ 
has been defined in \fr{definition-divergence}. 
In continuum mechanics a continuity equation for a time-dependent
  scalar quantity with an associated current leads to 
  a conservation law by integration over a spatial region. Correspondingly,  
the equivalent form \fr{cohomology-master-equations-divergence} of 
\fr{homology-master-equations} leads to
a conserved law by summing $p_{\,t}$ over all states
(See Remark\,\ref{remark-general-master-equation-conservation-law}).
Besides, in the study of master equations the form similar to 
\fr{cohomology-master-equations-divergence} is well-known 
in another formulation  
(See Refs.\,\cite{Polettini2012,Polettini2015} for example).  

Theorem\,\ref{fact-master-equations-current} 
deals with $p_{\,t}\in\Lambda^{\,0}(G)$. By contrast,  
an equality on $\Lambda^{\,1}(G)$  
associated with the master equations is obtained 
from \fr{cohomology-master-equations} as 
\beq
\frac{\dr}{\dr t}\dr p_{\,t}
=\dr\dr^{\dagger}\cI_{\,t}
=-\,\Delta_{\,E}\,\cI_{\,t},
\label{dp-master-equations}
\eeq
where $\Delta_{\,E}$ has been defined in \fr{Delta1}. 

  An example is shown below
  so that how introduced objects can be calculated from
  a given set of  master equations, and that how 
\fr{cohomology-master-equations} is applied.
  \begin{Example}
\label{example-one-way-ring}
    (A one-way interaction model on a ring).  
Consider the set of master equations whose total number of states is $N<\infty$,
\beq
\frac{\dr}{\dr t}p_{\,t}(x_{\,j})
=w_{\,j-1\to j}\,p_{\,t}(x_{\,j-1})
-\,w_{\,j\to j+1}\,p_{\,t}(x_{\,j}),\qquad j=0,\ldots,N-1,
\label{one-step-process-general}
\eeq
where
  $x_{\,N+j}=x_{\,j}$ so that $p_{\,t}(x_{\,N+j})=p_{\,t}(x_{\,j})$
for all $t\in \mbbR$, and $w_{j\to j+1}$ is a positive constant for each $j$. 
This model expresses a class of the so-called 
one-step processes \cite{VanKampen2007},
and can be  represented as the graph
$G^{\,\prime}=(V^{\,\prime},E^{\,\prime})$ depicted as follows:
$$
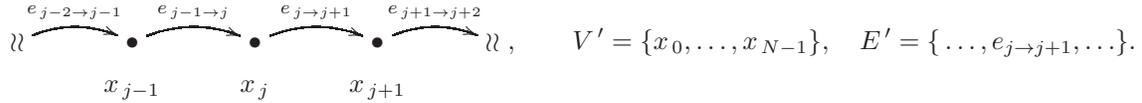
\xymatrix@R=2pt{
  \wr\wr\ar@/^/[r]^{e_{\,j-2\to j-1}}&\bullet\ar@/^/[r]^{e_{\,j-1\to j}}
  &\bullet\ar@/^/[r]^{e_{\,j\to j+1}}&\bullet\ar@/^/[r]^{e_{\,j+1\to j+2}}
  &\wr\wr\\
&x_{\,j-1}&x_{\,j}&x_{\,j+1}&
},\qquad
V^{\,\prime}=\{x_{\,0},\ldots,x_{\,N-1}\},\quad
E^{\,\prime}=\{\,\ldots, e_{j\to j+1},\ldots \}.
$$
In the graph above, there is no $\ol{e}$ for a given $e\in E^{\,\prime}$.
Introducing the edge set $E$ from $E^{\,\prime}$
such that there exists $\ol{e}\in E$ for all $e\in E$, one has the graph 
$G=(V,E)$ with $V=V^{\,\prime}$ depicted as 
$$
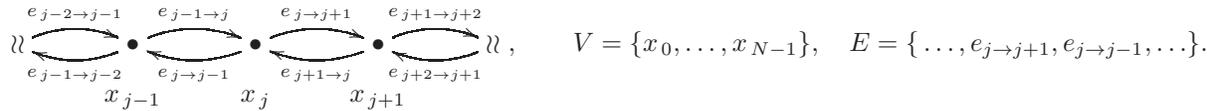
\xymatrix@R=5pt{
  \wr\wr\ar@/^/[r]^{e_{\,j-2\to j-1}}
  &\bullet\ar@/^/[r]^{e_{\,j-1\to j}}\ar@/^/[l]^{e_{\,j-1\to j-2}}
  &\bullet\ar@/^/[r]^{e_{\,j\to j+1}}\ar@/^/[l]^{e_{\,j\to j-1}}
  &\bullet\ar@/^/[r]^{e_{\,j+1\to j+2}}\ar@/^/[l]^{e_{\,j+1\to j}}
  &\wr\wr\ar@/^/[l]^{e_{\,j+2\to j+1}}\\
&x_{\,j-1}&x_{\,j}&x_{\,j+1}&
},\qquad
V=\{x_{\,0},\ldots,x_{\,N-1}\},\quad
E=\{\,\ldots, e_{j\to j+1}, e_{j\to j-1},\ldots \}.
$$
  To gain physical insight by analyzing a simple case,
  all the constants $\{w_{\,j-1\to j}\}$ are assumed to
  be equal, which is denoted by $\mu>0$. Then, to simplify the
  notation, $p_{\,t}(x_{\,j})$ is abbreviated to $p_{\,t}(j)$ so that
\fr{one-step-process-general} is written as 
  $$
  \frac{\dr}{\dr t}p_{\,t}(j)
  =\mu(\,p_{\,t}(j-1)-p_{\,t}(j)\,),\qquad j=0,\ldots,N-1.
  $$
Introduce $w\in C^{\,1}(G)$ such that 
$$
w(\,e_{\,j\to j+1}\,)
=\mu 
\quad\mbox{and}\quad
w(\,\ol{e_{\,j\to j+1}}\,)
=w(\,e_{\,j+1\to j}\,)
=0,\qquad
  j=0,\ldots,N-1.
$$ 
From \fr{cohomology-master-equations-divergence}, 
the divergence of $\cI_{\,t}$ at $x_{\,j}$ is expressed as 
$$
(\,\ddiv\,\cI_{\,t}\,)(j)
=-\,\mu(\,p_{\,t}(j-1)-p_{\,t}(j)\,)
,\qquad
j=0,\ldots,N-1.
$$
For completeness,
the solution satisfying the conservation law 
$\sum_{j=0}^{N-1}p_{\,t}(x_{\,j})=1$ for all $t\in\mbbR$ is shown.
First, introduce the Fourier transform of $p_{\,t}(j)$ with respect to
$j$ as 
$$
\wh{p}_{\,t}(k)
:=\frac{1}{N}\sum_{j=0}^{N-1}\e^{\,2\pi \imath jk/N}\,p_{\,t}(j),\qquad
k=0,\ldots,N-1.
$$
From this expression of $\wh{p}_{\,t}(k)$, it is straightforward to show that 
$$
\wh{p}_{\,t}(0)
=\frac{1}{N}\sum_{l=0}^{N-1}p_{\,t}(l)\ \in\mbbR,\qquad\mbox{and}\qquad
p_{\,t}(j)
=\sum_{k=0}^{N-1}\e^{\,-2\pi\imath jk/N}\ \wh{p}_{\,t}(k),
\qquad
j=0,\ldots,N-1.
$$
Second, it follows that
the Fourier transformed variables $\wh{p}_{\,t}(k)$ satisfy 
$$
\frac{\dr}{\dr t}\wh{p}_{\,t}(k)
=\mu(\,\e^{\,2\pi\imath k/N}-1\,)\,\wh{p}_{\,t}(k),\qquad
k=0,\ldots,N-1.
$$
The solution to its initial value problem is immediately obtained as
$$
\wh{p}_{\,t}(k)
=\exp\left[\,\mu\,t\,
  \{\,(\cos(2\pi k/N)-1)+\imath \sin(2\pi k/N)\,\}\,\right]\,
\wh{p}_{\,0}(k),\qquad
k=0,\ldots,N-1,
$$
which contains the conservation law at $k=0$, 
$$
\wh{p}_{\,t}(0)
=\wh{p}_{\,0}(0),\qquad\mbox{i.e.,}\qquad
\sum_{l=0}^{N-1}p_{\,t}(l)
=\sum_{l=0}^{N-1}p_{\,0}(l).
$$
Finally, the solution to the initial value problem in terms of the
original variables is 
$$
p_{\,t}(j)
=\sum_{k=0}^{N-1}\e^{\,-2\pi\imath jk/N}\ \e^{\ \mu\,t\,
  \{\,(\cos(2\pi k/N)-1)+\imath \sin(2\pi k/N)\,\}}\,\wh{p}_{\,t}(k).
$$
  \end{Example}

  %
\subsubsection{Expectation values}
\label{section-expectation-values-general-w}
Much attention is devoted to expectation values with respect to 
$p_{\,t}$ in applications of master equations. 
To discuss such expectation values, one introduces $\cO^{\,0}\in\Lambda^{\,0}(G)$ 
that is to be summed over vertexes at fixed time.
In this subsection, $m_{\,V}\in\Lambda^{\,0}(G)$ and 
$m_{\,E}\in  S^{\,1}(G)$ are chosen as \fr{trivial-measures}.

The expectation value of $\cO^{\,0}$ with respect to $p_{\,t}$ 
is denoted by 
\beq
\mbbE_{\,p_{t}}[\,\cO^{\,0}\,]
:=\sum_{x\in V}p_{\,t}(x)\,\cO^{\,0}(x),
\label{expectation-value-general-p-O}
\eeq
which is written in terms of the inner product with \fr{trivial-measures}
as 
$$
\mbbE_{\,p_{t}}[\,\cO^{\,0}\,]
=\inp{p_{\,t}}{\cO^{\,0}}_{\,V}.
$$   

From the master equations \fr{cohomology-master-equations}  
and Lemma\,\ref{Fact-d-d-dagger-1},  
the time-development of
$$
\avgg{\cO^{\,0}}{V}
:=\mbbE_{\,p_{t}}[\,\cO^{\,0}\,],
$$
is described by 
\beq
\frac{\dr}{\dr t}\avgg{\cO^{\,0}}{V}
=\inp{\dr^{\dagger}{\cI_{\,t}}}{\cO^{\,0}}_{\,V}
=\inp{\cI_{\,t}}{\dr\cO^{\,0}}_{\,E}. 
\label{average-time-development-vertex-1}
\eeq 
Equivalently, it follows from 
\fr{cohomology-master-equations-divergence} that
$$
\frac{\dr}{\dr t}\inp{p_{\,t}}{\cO^{\,0}}_{\,V}
+\inp{\ddiv \cI_{\,t}}{\cO^{\,0}}_{\,V}
=0.
$$
    
\begin{Remark}
\label{remark-general-master-equation-conservation-law}
Combining the sum over vertexes at fixed time $t$,
$$
\mbbE_{\,p_{t}}\,[\,1_{\,V}\,]
=\inp{p_{\,t}}{1_{\,V}}_{\,V}
=\sum_{x\in V}p_{\,t}(x),
$$
the identity 
$$
(\dr 1_{\,V})(e)
=1_{\,V}(\rmt{e})-1_{\,V}(\rmo{e})
=1-1
=0,
$$
and \fr{average-time-development-vertex-1}, one verifies that 
the derivative of $\avgg{1_{\,V}}{V}=\mbbE_{\,p_{t}}\,[\,1_{\,V}\,]$ 
with respect to time vanishes: 
$$
\frac{\dr}{\dr t}
\mbbE_{\,p_{t}}\,[\,1_{\,V}\,]
=\frac{\dr}{\dr t}\avgg{1_{\,V}}{V}
=\inp{\dot{p}_{\,t}}{1_{\,V}}_{\,V}
=\inp{\cI_{\,t}}{\dr 1_{\,V}}_{\,V}
=0,
$$
where $\dot{p}_{\,t}$ denotes derivative of 
$p_{\,t}$ with respect to time $t$, 
$\dot{p}_{\,t}=\dr p_{\,t}/\dr t$.
\end{Remark}
\begin{Remark}
For $\cO^{\,\prime}=\cO+c\,1_{\,V}\in\Lambda^{\,0}(G)$ with $c\in\mbbR$ 
being constant, one has
$$
\frac{\dr}{\dr t}\avgg{\cO^{\,\prime}}{V}
=\inp{\cI_{\,t}}{\dr \cO^{\,\prime}}_{\,E}
=\frac{\dr}{\dr t}\avgg{\cO}{V}.
$$
\end{Remark}

The inner product for $\cO_{\,A}^{\,1}\in \Lambda^{\,1}(G)$ and 
$\cO_{\,B}^{\,1}\in \Lambda^{\,1}(G)$ is $\inp{\cO_{\,A}^{\,1}}{\cO_{\,B}^{\,1}}_{\,E}$.
In the case where $\cO_{\,B}^{\,1}=\dr \dot{p}_{\,t}$  
and $\cO_{\,A}^{\,1}$ 
that does not depend on time $t$, one uses 
\fr{dp-master-equations} to obtain  
$$
\frac{\dr}{\dr t}\inp{\cO_{\,A}^{\,1}}{\dr p_{\,t}}_{\,E}
=\inp{\cO_{\,A}^{\,1}}{\dr \dot{p}_{\,t}}_{\,E}
=-\,\inp{\cO_{\,A}^{\,1}}{\Delta_{\,E}\,\cI_{\,t}}_{\,E}.
$$

\subsection{Detailed balance conditions}
Let $p^{\,\eq}$ be a prescribed probability distribution function so that 
$$
\sum_{x}p^{\,\eq}(x)
=1.
$$
Impose for any connected states $x$ and $y$
\beq
p^{\,\eq}(x)\,w_{\,x\to y}
=p^{\,\eq}(y)\,w_{\,y\to x},
\label{impose-detailed-balance-conditions-0}
\eeq
which are known as the
{\it detailed balance conditions}\,\cite{VanKampen2007,Schnakenberg1976}. 
These conditions imply the following. 
When $p^{\,\eq}$ is a  stationary solution 
of the master equations, 
$\dr p^{\,\eq}(x)/\dr t=0$,  one has from 
\fr{standard-master-equations} that 
\beq
\sum_{x'}\left[\,w_{\,x\to x'}p^{\,\eq}(x)
-w_{\,x'\to x}\,p^{\,\eq}(x^{\,\prime})\,\right]
=0.
\label{impose-detailed-balance-conditions-1}
\eeq
If \fr{impose-detailed-balance-conditions-0} holds, then 
\fr{impose-detailed-balance-conditions-1} is satisfied.
In this subsection \fr{impose-detailed-balance-conditions-0} is assumed
to hold.

These conditions are
written in the graph theoretic language as 
$$
\sum_{x\in V}p^{\,\eq}(x)
=1,
$$
and
\beq
p^{\,\eq}(\rmo{e})\,w(e)
=p^{\,\eq}(\rmt{e})\,w(\,\ol{e}\,). 
\label{detailed-balance-graph}
\eeq
The later leads to 
$$
w(\,\ol{e}\,)
=\pi(e)\,w(e),\qquad\mbox{where}\qquad 
\pi(e)
:=\frac{p^{\,\eq}(\rmo{e})}{p^{\,\eq}(\rmt{e})}
=\frac{1}{\pi(\,\ol{e}\,)},
$$
if $p^{\,\eq}(x)$ does not vanish for all $x\in V$.  
Thus, $\pi(e)=1$ for any loop edge $e\in E$, and
$$
\pi\in C_{\,\rmR}^{\,1}(G).
$$
To seek a reversible measure incorporating \fr{detailed-balance-graph}, let $m_{\,E}\in S^{\,1}(G)$ be such that 
$$
m_{\,E}(e)
=p^{\,\eq}(\rmo{e})\,w(e),
$$
from which   
$m_{\,E}(e)=m_{\,E}(\,\ol{e}\,) >0$. 
Thus, this 
$m_{\,E}\in S^{\,1}(G)$ can be used for 
a reversible measure defining \fr{inner-product-edge} 
if $m_{\,E}$ does not vanish.

The master equations \fr{homology-master-equations} under the 
detailed balance conditions are written as 
\beq
\frac{\dr}{\dr t}p_{\,t}(x)
=\sum_{e\in E_{\,x}}w(e)\left[\,
\pi(e)\,p_{\,t}(\rmt{e})-p_{\,t}(\rmo{e})\,\right].
\label{detailed-balance-master-equation}
\eeq
This can be written in terms of $\dr_{\pi}$, that is \fr{d-varphi} with 
$\varphi=\pi$, as 
\beq
\frac{\dr}{\dr t}p_{\,t}(x)
=\sum_{e\in E_{\,x}}w(e)\left[\,
(\dr_{\pi}\,p_{\,t})\,(e)\,\right]
=-\sum_{e\in E_{\,x}}w(\,\ol{e}\,)\left[\,(\dr_{\pi} p_{\,t})(\ol{e})\,\right].
\label{detailed-balance-master-equation-d-phi}
\eeq
The above equations involve $\dr_{\,\pi}$. 
In \fr{detailed-balance-master-equation-d-phi}, there is no 
loop edge contribution   
since $\pi(e)=1$ and $(\dr_{\,\pi}\, p_{\,t})(e)=0$ for any loop edge $e\in E$, 
(See Remark\,\ref{fact-d-phi-becomes-the-standard-d}). 
Although there is no self-adjoint 
operator in \fr{detailed-balance-master-equation-d-phi},  
there exists a way to express master equations in terms of a Laplacian, 
which is accomplished by a change of variables.

The following is the main theorem in this paper:
\begin{Theorem}
\label{fact-master-equations-detailed-balance}
(Diffusion equations from master equations). 
Consider 
the master equation
\fr{homology-master-equations} in  
the case that 
\begin{enumerate}
\item
  the prescribed stationary solution does not vanish, that is, $p^{\,\eq}(x)\neq 0$,
  for all $x\in V$, and 
\item 
the detailed balance conditions are satisfied, 
that is, $w(e)$ satisfies \fr{detailed-balance-graph}.
\end{enumerate}  
Choose the measures $m_{\,V}\in\Lambda^{\,0}(G)$ and $m_{\,E}\in S^{\,1}(E)$ 
to be 
\beq
m_{\,V}(x)
=p^{\,\eq}(x),\qquad\mbox{and}\qquad 
m_{\,E}(e)
=w(e)\,p^{\,\eq}(\rmo{e}), 
\label{detailed-balance-measures}
\eeq   
respectively.  
Then introduce $\psi_{\,t}\in\Lambda^{\,0}(G)$ such that 
\beq
p_{\,t}(x)
=p^{\,\eq}(x)\,\psi_{\,t}(x),\qquad \forall\,x\in V,
\label{detailed-balance-split-wave-function}
\eeq
where $p_{\,t}$ is a solution to the master equation
\fr{homology-master-equations}. 
This function $\psi_{\,t}$ satisfies the diffusion
equation
\beq
\frac{\dr}{\dr  t}\psi_{\,t}
=\Delta_{\,V}\,\psi_{\,t},
\label{detailed-balance-diffusion-equation}
\eeq
where $\Delta_{\,V}$ has been defined in \fr{Delta0}.  
\end{Theorem}
\begin{Proof}
Substituting  \fr{detailed-balance-split-wave-function}
into \fr{detailed-balance-master-equation}, and using 
the definition of the Laplacian \fr{Delta0}, one can complete the 
proof. The details are as follows:

From \fr{detailed-balance-split-wave-function}
and \fr{detailed-balance-master-equation}, and $\dr p^{\,\eq}(x)/\dr t=0$,
one has 
\beqa
p^{\,\eq}(x)\frac{\dr}{\dr t}\psi_{\,t}\,(x)
&=&\sum_{e\in E_{\,x}}w(e)\,\left[\,
\frac{p^{\,\eq}(\,\rmo{e}\,)}{p^{\,\eq}(\,\rmt{e}\,)}\,
p^{\,\eq}(\,\rmt{e}\,)\,\psi_{\,t}(\,\rmt{e}\,)
-p^{\,\eq}(\,\rmo{e}\,)\,\psi_{\,t}(\,\rmo{e}\,)
\,\right]
\non\\
&=&
\sum_{e\in E_{\,x}}w(e)\,\left[\,
p^{\,\eq}(\,\rmo{e}\,)\,\psi_{\,t}(\,\rmt{e}\,)
-p^{\,\eq}(\,\rmo{e}\,)\,\psi_{\,t}(\,\rmo{e}\,)
\,\right],
\non
\eeqa
from which 
$$
\frac{\dr}{\dr t}\psi_{\,t}\,(x)
=\frac{1}{p^{\,\eq}(x)}
\sum_{e\in E_{\,x}}w(e)\,p^{\,\eq}(\,\rmo{e}\,)\,
\left[\,
\psi_{\,t}(\,\rmt{e}\,)-\,\psi_{\,t}(\,\rmo{e}\,)
\,\right].
$$
Since the action of the Laplacian \fr{Delta0-explicit-form}
with \fr{detailed-balance-measures} on $\psi_{\,t}$ is  
\beq
(\,\Delta_{\,V}\,\psi_{\,t}\,)(x)
=\frac{1}{p^{\,\eq}(x)}\sum_{e\in E_{\,x}}
p^{\,\eq}(\,\rmo{e}\,)\,w(e)
\left[\,\psi_{\,t}(\,\rmt{e}\,)-\,\psi_{\,t}(\,\rmo{e}\,)
\,\right],
\label{Delta0-explicit-form-detailed-balance-non-reduced}
\eeq
one obtains \fr{detailed-balance-diffusion-equation}. 
\qed
\end{Proof}

\begin{Remark}
The explicit expression \fr{Delta0-explicit-form-detailed-balance-non-reduced} 
reduces further.
It follows from 
$\rmo{e}=x$ for $e\in E_{\,x}$ that 
\beq
(\,\Delta_{\,V}\,\psi_{\,t}\,)(x)
=\sum_{e\in E_{\,x}}
\,w(e)
\left[\,\psi_{\,t}(\,\rmt{e}\,)-\,\psi_{\,t}(\,\rmo{e}\,)
\,\right]
=\sum_{e\in E_{\,x}}
\,w(e)\,(\dr\psi_{\,t})(e).
\label{Delta0-explicit-form-detailed-balance-reduced}
\eeq 
\end{Remark}

\begin{Remark}
The conservation law of the sum 
$$
\sum_{x\in V}p_{\,t}(x)
=1,\qquad \forall t\in\mbbR
$$
can be shown below. With \fr{detailed-balance-diffusion-equation}, 
one can show
\beqa
\frac{\dr}{\dr t}\sum_{x\in V}p_{\,t}(x)
&=&\sum_{x\in V}p^{\,\eq}(x)\left(\frac{\dr}{\dr t}\psi_{\,t}\right)(x)
=\sum_{x\in V}p^{\,\eq}(x)\left(\Delta_{\,V}\psi_{\,t}\right)(x)
\non\\
&=&\inp{\Delta_{\,V}\psi_{\,t}}{1_{\,V}}_{\,V}
=\inp{\psi_{\,t}}{\Delta_{\,V}1_{\,V}}_{\,V}
=0.
\label{detailed-balance-sum-p-over-vertex}
\eeqa
\end{Remark}

\begin{Remark}
\label{fact-master-equations-detailed-balance-no-loop}
Theorem\,\ref{fact-master-equations-detailed-balance} 
and \fr{Delta0-explicit-form}   
indicate that 
\fr{detailed-balance-diffusion-equation}  
does not involve any loop edge.  
This property also holds for the equations written in terms of 
$p_{\,t}(x)$, \fr{detailed-balance-master-equation}.
\end{Remark}

\begin{Remark}
Since the diffusion equation
\fr{detailed-balance-diffusion-equation} can be written as 
$$
\frac{\dr}{\dr t}\psi_{\,t}
+\,\dr^{\dagger}\dr\,\psi_{\,t}
=0,
$$
one has the form of continuity equation by introducing
$\Pi_{\,t}\in\Lambda^{\,1}(G)$  
\beq
\frac{\dr}{\dr t}\psi_{\,t}+\ddiv\, \Pi_{\,t}
=0,\qquad
\Pi_{\,t}
:=-\,\dr\,\psi_{\,t},
\label{cohomology-master-equations-divergence-psi}
\eeq
where $\ddiv:\Lambda^{\,1}(G)\to \Lambda^{\,0}(G)$ 
has been defined in \fr{definition-divergence}.
For the case where the detailed balance conditions need not hold, but 
the conditions $w(\ol{e})=w(e)$ hold for all $e\in E$,
Fick's law \fr{Fick-law-general} holds.
Besides, in the present case, the corresponding Fick's law is
$$
\Pi_{\,t}(e)
=-\,(\dr\,\psi_{\,t})(e)
$$
for all $e\in E$. Hence the current on $e\in E$
is expressed as the difference between $\psi_{\,t}(\rmt{e})$ and
$\psi_{\,t}(\rmo{e})$ as follows: 
$$
\Pi_{\,t}(e)=-\psi_{\,t}(\rmt{e})+\psi_{\,t}(\rmo{e}).
$$
  \end{Remark}    

The following is an example of how this formulation can be applied to 
a model studied in nonequilibrium statistical mechanics: 
\begin{Example}
\label{example-kinetic-ising-psi}
(Kinetic Ising model without spin-coupling, \cite{Goto2015}).  
Let $\sigma=\pm 1$ be a spin variable, and 
$p_{\,\Ising}^{\,\eq}(\sigma)$ an equilibrium distribution of $\sigma$.
Consider the master equations
$$
\frac{\dr}{\dr t}p_{\,t,\Ising}(\sigma)
=-w_{\,\sigma\to-\,\sigma}\,p_{\,t,\Ising}(\sigma)
+w_{\,-\sigma\to \sigma}\,p_{\,t,\Ising}(-\sigma),\qquad \sigma=\pm1,
$$ 
where the detailed balance condition is satisfied: 
$$
w_{\,\sigma\to-\,\sigma}\,p_{\,\Ising}^{\,\eq}(\sigma)
=w_{\,-\sigma\to\sigma}\,p_{\,\Ising}^{\,\eq}(-\sigma).
$$ 
This set of master equations 
induces 
$G=(V,E)$ and
  $w\in C^{\,1}(G)$. This $G$ consists of $V=\{1,-1\}$ and 
  $E=\{e^{\,\prime},e^{\,\prime\prime}\}$, where $E_{\,1}=\{e^{\,\prime}\}$ and
  $E_{\,-\,1}=\{e^{\,\prime\prime}\}$.
  The $G$ and $w\in C^{\,1}(G)$ are such that 
$$
\xymatrix@R=2pt{
&\bullet\ar@/^/[r]^{e^{\,\prime}}&\bullet\ar@/^/[l]^{e^{\,\prime\prime}}\\
&\sigma=1&\sigma=-1
},
\quad
w(\,e^{\,\prime}\,)
=w_{\,1\to -\,1},\quad
w(\,e^{\,\prime\prime}\,)
=w_{\,-\,1\to\, 1},
$$
Introduce $\psi_{\,t,\Ising}$ such that 
$$
p_{\,t,\Ising}(\sigma)
=p_{\,\Ising}^{\,\eq}(\sigma)\psi_{\,t,\Ising}(\sigma),\qquad
\sigma=\pm1.
$$
With these variables and the detailed balance condition  
$$
w_{\,-\,\sigma\to\sigma}
=w_{\,\sigma\to-\,\sigma}\,
\frac{p_{\,\Ising}^{\,\eq}(\sigma)}{p_{\,\Ising}^{\,\eq}(-\sigma)},
$$
one has that 
$$
p_{\,\Ising}^{\,\eq}(\sigma)
\frac{\dr}{\dr t}\psi_{\,t,\Ising}(\sigma)
=\,w_{\,\sigma\to-\,\sigma}\,\left[\,-\,
p_{\,\Ising}^{\,\eq}(\sigma)\psi_{\,t,\Ising}(\sigma)
+
\frac{p_{\,\Ising}^{\,\eq}(\sigma)}{p_{\,\Ising}^{\,\eq}(-\sigma)}p_{\,\Ising}^{\,\eq}(-\sigma)\psi_{\,t,\Ising}(-\sigma)\right],
$$
from which 
\beq
\frac{\dr}{\dr t}\psi_{\,t,\Ising}(\sigma)
=\,w_{\,\sigma\to-\,\sigma}\,\left[\,\psi_{\,t,\Ising}(-\sigma)
-\,\psi_{\,t,\Ising}(\sigma)\,\right],\qquad \sigma=\pm1.
\label{psi-kinetic-Ising}
\eeq
This derived set of equations is consistent with 
\fr{Delta0-explicit-form-detailed-balance-reduced}. 
\end{Example}
      Note that diffusion equations in 
      Example\,\ref{example-one-way-ring} are not obtained by applying 
      Theorem\,\ref{fact-master-equations-detailed-balance}, since 
      the detailed balance conditions are not satisfied. 

To investigate the long-time behavior of the system 
  \fr{detailed-balance-diffusion-equation}, rewrite
\fr{detailed-balance-diffusion-equation} as   
\beq
\frac{\dr}{\dr t}\psi_{\,t}
=\cF(\psi_{\,t}),
\label{detailed-balance-diffusion-equation-F}
\eeq
where $\cF:\Lambda^{\,0}(G)\to \Lambda^{\,0}(G)$ has been defined such that 
$$
\cF(\psi)
:=\Delta_{\,V}\psi,\ \in \Lambda^{\,0}(G).
$$
The following Lemmas will be used:
\begin{Lemma}
\label{fact-laplace-equation-Vertex}
The non-trivial solutions, $\psi^{\,(0)}(x)\neq 0$,  
to the equations 
$(\Delta_{\,V}\psi^{\,(0)})(x)=0$ for any $x\in V$ 
are   
$\psi^{\,(0)}(x)=\psi_{\,0}^{\,(0)}$, where 
$\psi_{\,0}^{\,(0)}\in\mbbR$ is constant.
\end{Lemma} 
\begin{Proof}
For general $\psi$, it follows that
$$
\inp{\Delta_{\,V}\,\psi}{\psi}_{\,V}
=-\inp{\dr\psi}{\dr\psi}_{\,E}
\leq0.
$$
The equality holds only when $\dr \psi=0$. 
It implies that 
$$
\Delta_{\,V}\psi^{\,(0)}
=0\qquad
\Longleftrightarrow 
\qquad
\dr\psi^{\,(0)}
=0.
$$
Since 
$$
(\,\dr\psi^{\,(0)}\,)(e)
=\psi^{\,(0)}(\rmt{e})-\psi^{\,(0)}(\rmo{e})
=0,\qquad\forall e\in E,
$$
and the assumption that the graph is connected, 
the solution is $\psi^{\,(0)}(x)=\psi_{\,0}^{\,(0)}$  
with $\psi_{\,0}^{\,(0)}$ being constant. 
\qed
\end{Proof}
Lemma\,\ref{fact-laplace-equation-Vertex} states that 
the steady state $\psi^{\,(0)}$ for \fr{detailed-balance-diffusion-equation}, 
$\dr \psi^{\,(0)}/\dr t=0$, is $\psi^{\,(0)}=\psi_{\,0}^{\,(0)}1_{\,V}$.
In other words, this $\psi_{\,0}^{\,(0)}1_{\,V}$ 
forms a fixed point set 
for this system.  
The following describes the stability for 
$\psi=\psi_{\,0}^{\,(0)}1_{\,V}$: 

\begin{Lemma}
\label{fact-laplace-equation-Lyapunov}
The state $\psi_{\,t}=\psi_{\,0}^{\,(0)}1_{\,V}$ 
is asymptotically stable for 
\fr{detailed-balance-diffusion-equation-F},
where $\psi_{\,0}^{\,(0)}\in\mbbR$ is constant.  
\end{Lemma}
\begin{Proof}
Introduce the dynamical system for $\wt{\psi}\in\Lambda^{\,0}(G)$,
\beq
\frac{\dr}{\dr t}\wt{\psi}_{\,t}
=\wt{\cF}\left(\,\wt{\psi}_{\,t}\,\right),
\label{Lyapunov-stability-shifted-dynamical-system}
\eeq
where $\wt{\cF}:\Lambda^{\,0}(G)\to \Lambda^{\,0}(G)$ has been defined such that 
$$
\wt{\cF}\left(\,\wt{\psi}_{\,t}\,\right)
:=\Delta_{\,V}\left(\,\wt{\psi}_{\,t}+{\psi_{\,0}^{\,(0)}}1_{\,V}\,\right).
$$
It immediately follows 
 that  
$$
\wt{\cF}(0_{\,V})
=\cF(\,\psi_{\,0}^{\,(0)}1_{\,V}\,)
=\Delta_{\,V}(\,\psi_{\,0}^{\,(0)}1_{\,V}\,)
=0_{\,V},
$$
where 
$0_{\,V}$ is the element of $\Lambda^{\,0}(G)$ such that $0_{\,V}(x)=0$ for 
any $x\in V$. 
This states that 
$0_{\,V}\in \Lambda^{\,0}(G)$ is 
a fixed point set for 
\fr{Lyapunov-stability-shifted-dynamical-system}.
The relation between \fr{Lyapunov-stability-shifted-dynamical-system}
and \fr{detailed-balance-diffusion-equation-F} is as follows.
If $\psi_{\,t}$  is a solution to \fr{detailed-balance-diffusion-equation-F}, 
then  $\wt{\psi}_{\,t}=\psi_{\,t}-\psi_{\,0}^{\,(0)}\,1_{\,V}$ is a solution to
\fr{Lyapunov-stability-shifted-dynamical-system}. This is due to 
$$
\frac{\dr}{\dr t}\wt{\psi}_{\,t}
=\frac{\dr}{\dr t}(\psi_{\,t}-\psi_{\,0}^{\,(0)}\,1_{\,V})
=\frac{\dr}{\dr t}\psi_{\,t}
=\cF(\psi_{\,t})
=\Delta_{\,V}\,\psi_{\,t}
=\Delta_{\,V}\,\left(\wt{\psi}_{\,t}+\psi_{\,0}^{\,(0)}1_{\,V}\right)
=\wt{\cF}\left(\wt{\psi}_{\,t}\right).
$$
On the contrary, if $\wt{\psi}_{\,t}$  is a solution to
\fr{Lyapunov-stability-shifted-dynamical-system}, 
then  $\psi_{\,t}=\wt{\psi}_{\,t}+\psi_{\,0}^{\,(0)}\,1_{\,V}$ is a solution to
\fr{detailed-balance-diffusion-equation-F}. This is due to
$$
\frac{\dr}{\dr t}\psi_{\,t}
=\frac{\dr}{\dr t}(\wt{\psi}_{\,t}+\psi_{\,0}^{\,(0)}\,1_{\,V})
=\frac{\dr}{\dr t}\wt{\psi}_{\,t}
=\wt{\cF}\left(\wt{\psi}_{\,t}\right)
=\Delta_{\,V}\,\left(\wt{\psi}_{\,t}+\psi_{\,0}^{\,(0)}1_{\,V}\right)
=\Delta_{\,V}\,\psi_{\,t}
=\cF(\psi_{\,t}).
$$
Hence, the solution to
  \fr{detailed-balance-diffusion-equation-F} is obtained by
  $\psi_{\,t}=\wt{\psi}_{\,t}+\psi_{\,0}^{\,(0)}1_{\,V}$ with $\wt{\psi}_{\,t}$
  being a solution to \fr{Lyapunov-stability-shifted-dynamical-system}.

Then define $\cL:\Lambda^{\,0}(G)\to\mbbR$  
such that 
$$
\cL\left(\,\wt{\psi}_{\,t}\,\right)
=\frac{1}{2}\inp{\wt{\psi}_{\,t}}{\wt{\psi}_{\,t}}_{\,V}.
$$
It  immediately follows that 
$\cL(\,\wt{\psi}_{\,t}\,)\geq 0$ for any $t\in\mbbR$,
where $\wt{\psi}_{\,t}$ is a solution to
  \fr{Lyapunov-stability-shifted-dynamical-system}.   
In addition, one has that  $\dr \cL(\wt{\psi}_{\,t})/\dr t\leq 0$ 
for any 
$t\in \mbbR$,  since
$$
\frac{\dr}{\dr t}\cL\left(\,\wt{\psi}_{\,t}\,\right)
=\inp{\dot{\wt{\psi}_{\,t}}}{\wt{\psi}_{\,t}}_{\,V}
=\inp{\Delta_{\,V}\left(\wt{\psi}_{\,t}+\psi_{\,0}^{\,(0)}1_{\,V}\right)}{
\wt{\psi}_{\,t}}_{\,V}
=-\inp{\dr^{\,\dagger}\dr\,\wt{\psi}_{\,t}}{\wt{\psi}_{\,t}}_{\,V}
=-\inp{\dr\wt{\psi}_{\,t}}{\dr\wt{\psi}_{\,t}}_{\,E}
\leq 0.
$$
From these properties, $\cL$ 
is a Lyapunov function. 
Applying these statements to the Lyapunov stability theorem, one has that 
$\wt{\psi}_{\,t}=0_{\,V}$ in the 
dynamical system 
\fr{Lyapunov-stability-shifted-dynamical-system} 
is asymptotically stable.
This yields that $\psi_{\,0}^{\,(0)}1_{\,V}$ is 
asymptotically stable for \fr{detailed-balance-diffusion-equation-F}.
\qed
\end{Proof}
Then one has the following from
Theorem\,\ref{fact-master-equations-detailed-balance}: 
\begin{Corollary}
(Convergence of solutions to master equations).
\label{fact-master-equations-detailed-balance-relaxation}
$$
\psi_{\,0}^{\,(0)}
  =1,\qquad\mbox{and}\qquad
\lim_{t\to\infty}p_{\,t}(x)
=p^{\,\eq}(x),\qquad\forall x\in V. 
$$
\end{Corollary}
\begin{Proof}
It follows from Lemmas\,\ref{fact-laplace-equation-Vertex} 
and \,\ref{fact-laplace-equation-Lyapunov} that   
$$
\lim_{t\to\infty}p_{\,t}(x)
=p^{\,\eq}(x)\lim_{t\to\infty}\psi_{\,t}(x)
=p^{\,\eq}(x)\,\psi_{\,0}^{\,(0)}.
$$
This and the normalization condition 
$\sum_{x\in V}p^{\,\eq}(x)=1$ lead to  
$$
\sum_{x\in V}\lim_{t\to\infty}p_{\,t}(x)
=\psi_{\,0}^{\,(0)}\,\sum_{x\in V}p^{\,\eq}(x)
=\psi_{\,0}^{\,(0)}.
$$
Meanwhile,  the conservation law 
\fr{detailed-balance-sum-p-over-vertex} holds. Thus, the left hand side 
of the equation above is unity: 
$$
\psi_{\,0}^{\,(0)}
=1.
$$
Therefore, 
$$
\lim_{t\to\infty}p_{\,t}(x)
=p^{\,\eq}(x).
$$
\qed
\end{Proof}

It has been shown that discrete diffusion equations 
are derived from master equations under the conditions that the 
detailed balance conditions are satisfied.
Its converse statement under these conditions also holds.
\begin{Proposition}
\label{fact-diffusion-yield-master}
(From diffusion equations to master equations).  
Let $w\in C^{\,1}(G)$ be a transition matrix, and $p^{\,\eq}\in\Lambda^{\,0}(G)$ 
an equilibrium distribution function. 
Assume that the detailed balance conditions \fr{detailed-balance-graph}
hold. 
Choose $m_{\,V}\in \Lambda^{\,0}(G)$ and $m_{\,E}\in S^{\,1}(G)$ so that  
$m_{\,V}(x)=p^{\,\eq}(x)$ and $m_{\,E}(e)=w(e)p^{\,\eq}(\rmo{e})$ as in 
\fr{detailed-balance-measures}.
Then, the diffusion equations \fr{detailed-balance-diffusion-equation}
yield  master equations.
\end{Proposition}
\begin{Proof}
Introduce the inner products \fr{inner-product-vertex} 
and \fr{inner-product-edge}. In addition, introduce  
$p_{\,t}\in\Lambda^{\,0}(G)$ that depends on $t\in\mbbR$ 
such that $p_{\,t}(x)=p^{\,\eq}(x)\psi_{\,t}(x)$.
Then multiplying both sides of the diffusion equations by $p^{\,\eq}(x)$,
one has
$$
p^{\,\eq}(x)\frac{\dr}{\dr t}\psi_{\,t}(x)
=\sum_{e\in E_{\,x}}p^{\,\eq}(\,\rmo{e}\,) w(e)\,
\left[\,\psi_{\,t}(\,\rmt{e}\,)-\psi_{\,t}(\,\rmo{e}\,)\,\right],
\qquad
\forall\ x\in V.
$$ 
With \fr{detailed-balance-graph}, the equations above can be written as 
$$
p^{\,\eq}(x)\frac{\dr}{\dr t}\psi_{\,t}(x)
=\sum_{e\in E_{\,x}}\left[\,
p^{\,\eq}(\,\rmt{e}\,) w(\ol{e})
\psi_{\,t}(\,\rmt{e}\,)
-p^{\,\eq}(\,\rmo{e}\,) w(e)\psi_{\,t}(\,\rmo{e}\,)\,\right].
$$ 
This can also be written in terms of $p_{\,t}(x)=p^{\,\eq}(x)\psi_{\,t}(x)$ as 
$$
\frac{\dr}{\dr t}p_{\,t}(x)
=\sum_{e\in E_{\,x}}\left[\,
-p_{\,t}(\,\rmo{e}\,) w(e)
+p_{\,t}(\,\rmt{e}\,) w(\ol{e})
\,\right].
$$
\qed
\end{Proof}

So far, basic statements of the system have been given.
These include asymptotic behavior of the diffusion equations, 
Lemma\,\ref{fact-laplace-equation-Lyapunov}.  
Furthermore, if the eigenvalues and eigenfunctions 
of the Laplacian $\Delta_{\,V}$ 
are known, then an explicit solution for the initial value problem 
of the diffusion equations  
\fr{detailed-balance-diffusion-equation} can be obtained.
Since the diffusion equations are linear,  
it is enough to consider the case where the solution space 
is a linear space of $\Lambda^{\,0}(G)$. 
Moreover, from Theorem\,\ref{fact-master-equations-detailed-balance}, one has 
the equation on $\Lambda^{\,1}(G)$ 
\beq
\frac{\dr}{\dr t}\dr\psi_{\,t}
=\Delta_{\,E}\,\dr\psi_{\,t}.
\label{detailed-balance-split-wave-function-edge}
\eeq
Since  \fr{detailed-balance-split-wave-function-edge} is linear,  
eigenvalues and eigenfunctions of $\Delta_{\,E}$ play roles.
Thus in what follows 
$\Lambda^{\,0}(G)$ and $\Lambda^{\,1}(G)$ denote linear spaces.
Then how to construct an explicit solution to the diffusion equations 
is argued below.

Let $A$ be a linear operator acting on a finite-dimensional vector 
space $\cU$, and   
$\lambda_{\,A}$ a non-zero number satisfying  
$$
A\,\phi
=\lambda_{\,A}\phi,
$$ 
where $\phi\in\cU$ is not $0$. Then  
$\lambda_{\,A}$ and $\phi$
are referred to as an 
{\it non-zero eigenvalue} and its associated or corresponding 
{\it eigenfunction}, respectively.
For a self-adjoint operator $A$, 
it is known that all the eigenvalues are real, $\lambda_{\,A}\in\mbbR$. 
In addition, 
if all the non-zero eigenvalues are positive, then $A$ is referred to as 
a {\it positive operator}.
If $A\,\phi=0$ with non-zero $\phi$, 
then $\phi$ is referred to as the {\it eigenfunction} associated with 
the {\it zero-eigenvalue}.

Since the Laplacian $\Delta_{\,V}$ is a linear operator acting on 
$\Lambda^{\,0}(G)$ and is 
self-adjoint as shown in 
Lemma\,\ref{fact-Delta0-self-adjoint}, 
one can discuss if $-\Delta_{\,V}$ is a positive operator or not. 
Then one has the following: 
\begin{Lemma}
\label{fact-negative-Laplacian-V-positive-operator}
The negative of the Laplacian acting on $\Lambda^{\,0}(G)$, 
$-\Delta_{\,V}$, is a positive operator.
\end{Lemma}
\begin{Proof}
Let $\wt{\lambda}_{\,V}^{\,(s)}$ be a non-zero eigenvalue labeled by $s$, 
and $\wt{\phi}_{\,V}^{\,(s)}$ an associated eigenfunction
of $-\Delta_{\,V}$ so that 
$-\Delta_{\,V}\wt{\phi}_{\,V}^{\,(s)}=\wt{\lambda}_{\,V}^{\,(s)}\wt{\phi}_{\,V}^{\,(s)}$.  
Then it follows that 
$$
\wt{\lambda}_{\,V}^{\,(s)}
\inp{\wt{\phi}_{\,V}^{\,(s)}}{\wt{\phi}_{\,V}^{\,(s)}}_{\,V}
=\inp{\wt{\lambda}_{\,V}^{\,(s)}\wt{\phi}_{\,V}^{\,(s)}}{\wt{\phi}_{\,V}^{\,(s)}}_{\,V}
=\inp{-\Delta_{\,V}\wt{\phi}_{\,V}^{\,(s)}}{\wt{\phi}_{\,V}^{\,(s)}}_{\,V}
=\inp{\dr\wt{\phi}_{\,V}^{\,(s)}}{\dr\wt{\phi}_{\,V}^{\,(s)}}_{\,E}
\geq\, 0,
$$
for all $\wt{\phi}_{\,V}^{\,(s)}\neq 0_{\,V}$. 
Applying the assumed conditions 
$\inp{\wt{\phi}_{\,V}^{\,(s)}}{\wt{\phi}_{\,V}^{\,(s)}}_{\,V}>0$
and $\wt{\lambda}_{\,V}^{\,(s)}\neq 0$ 
to the obtained inequality   
$\wt{\lambda}_{\,V}^{\,(s)}
\inp{\wt{\phi}_{\,V}^{\,(s)}}{\wt{\phi}_{\,V}^{\,(s)}}_{\,V}\geq\, 0$, 
one concludes that 
$\wt{\lambda}_{\,V}^{\,(s)}>0$.
\qed
\end{Proof}
\begin{Remark}
\label{fact-all-eigenvalues-Delta-V-positive}
Repeating similar arguments in the proof of 
Lemma\,\ref{fact-negative-Laplacian-V-positive-operator}, 
one has that all the  eigenvalues of $\Delta_{\,V}$ are negative or zero,
$$ 
\Delta_{\,V}\phi_{\,V}^{\,(s)}
=\lambda_{\,V}^{\,(s)}\phi_{\,V}^{\,(s)},\qquad
\lambda_{\,V}^{\,(s)}
\leq 0,
$$
where $\phi_{\,V}^{\,(s)}\in\Lambda^{\,0}(G)$ is the eigenfunction 
associated with $\lambda_{\,V}^{\,(s)}$ for $s\neq 0$.
For $s=0$,   
$\phi_{\,V}^{\,(0)}$ denotes the eigenfunction associated with the
zero-eigenvalue  
$\lambda_{\,V}^{\,(0)}=0$. 
\end{Remark}

Moreover, given a self-adjoint operator $A$  
acting on a linear space $\cU$
with an inner product $\inp{}{}_{\,\cU}$,   
it is known that $\cU$ has an orthonormal basis consisting of eigenvectors of $A$.
By applying this, one has the decomposition  
\beq
\psi_{\,t}
=\sum_{s}a^{\,(s)}(t)\phi_{\,V}^{\,(s)}
=\sum_{s\in \cN_{\,V}}a^{\,(s)}(t)\phi_{\,V}^{\,(s)}+a^{\,(0)}(t)
\phi_{\,V}^{\,(0)}, 
\qquad\mbox{with}\qquad 
\inp{\phi_{\,V}^{\,(s)}}{\phi_{\,V}^{\,(s^{\,\prime})}}_{\,V}
=\delta^{\,ss^{\,\prime}},
\label{decomposition-psi-eigen-basis}
\eeq
where $s$ denotes a label for eigenfunction, $\cN_{\,V}$ the totality of 
labels for non-zero eigenfunctions, 
$a^{\,(s)}$ the real-valued function labeled by $s$ of $t$,  
and 
$\delta^{\,ss^{\,\prime}}$ the Kronecker delta, giving unity if $s=s^{\,\prime}$ 
and zero otherwise.

The normalized eigenfunction associated with the zero-eigenvalue 
$\phi_{\,V}^{\,(0)}$ 
is obtained as follows: 
\begin{Lemma} 
\label{fact-normalized-zero-eigenfunction}
  The normalized zero-eigenfunction is $\phi_{\,V}^{\,(0)}=1_{\,V}$:     
$$
\Delta_{\,V}\phi_{\,V}^{\,(0)}
=0,\qquad\mbox{and}\qquad
\inp{\phi_{\,V}^{\,(0)}}{\phi_{\,V}^{\,(0)}}_{\,V}
=1.
$$
\end{Lemma}
\begin{Proof}
It follows that $\Delta_{\,V}1_{\,V}=-\dr^{\,\dagger}\dr 1_{\,V}=0$. Then,   
the normalization condition is verified as  
$$
\inp{\phi_{\,V}^{\,(0)}}{\phi_{\,V}^{\,(0)}}_{\,V}
=\inp{1_{\,V}}{1_{\,V}}_{\,V}
=\sum_{x\in V}p^{\,\eq}(x)
=1.
$$ 
\qed
\end{Proof}
To discuss properties of the zero-eigenfunction further,
let $\ker\Delta_{\,V}$ be 
the space spanned by the zero-eigenfunction $\phi_{\,V}^{\,(0)}$  : 
$$
\ker\Delta_{\,V}
:=\{\ \phi_{\,V}\in\Lambda^{\,0}(G)\ |\ \Delta_{\,V}\,\phi_{\,V}=0 \ \}.
$$
Then, one has the following:
\begin{Lemma}
\label{fact-normalized-zero-eigenfunction-1-dimension-unique}  
A normalized zero-eigenfunction in 
Lemma\,\ref{fact-normalized-zero-eigenfunction} is unique 
up to sign, and 
$\dim(\ker\Delta_{\,V}^{\,(0)})=1$. 
\end{Lemma}
\begin{Proof}
First, $\dim(\ker\Delta_{\,V}^{\,(0)})=1$ is proved. 
The proof below is similar to the proof of
Lemma\,\ref{fact-laplace-equation-Vertex}.

For general $\psi\in \Lambda^{\,0}(G)$, it follows that
$$
\inp{\Delta_{\,V}\,\psi}{\psi}_{\,V}
=-\inp{\dr\,\psi}{\dr\,\psi}_{\,E}
\leq 0.
$$
The equality holds only when $\dr\,\psi=0$. It implies for
$\phi_{\,V}^{\,(0)}\in\ker\Delta_{\,V}$ that
$$
\Delta_{\,V}\phi_{\,V}^{\,(0)}
=0
\quad\Longleftrightarrow\quad
\dr\,\phi_{\,V}^{\,(0)}
=0.
$$
Thus, for all $e\in E$, 
$$
(\dr\,\phi_{\,V}^{\,(0)})(e)
=\phi_{\,V}^{\,(0)}(\rmt{e})-\phi_{\,V}^{\,(0)}(\rmo{e})
=0,
$$
or equivalently,
$$
\phi_{\,V}^{\,(0)}(\rmt{e})
=\phi_{\,V}^{\,(0)}(\rmo{e}).
$$
Combining this equality and the assumption that
the graph $G$ is connected, one has that
$$
\phi_{\,V}^{\,(0)}
=\phi_{\,V,0}^{\,(0)}\,1_{\,V},
$$
where $\phi_{\,V,0}^{\,(0)}\in\mbbR$ is constant. Thus
$\ker\Delta_{\,V}^{\,(0)}=\spanmath\{\,1_{\,V}\}$, and 
the equality 
$\dim(\ker\Delta_{\,V}^{\,(0)})=1$ holds.

In general, given a non-zero vector, the normalized vector
is uniquely determined up to sign.  
It follows from this uniqueness with $\dim(\ker\Delta_{\,V}^{\,(0)})=1$ 
that the normalized zero-eigenfunction is
uniquely determined up to sign. 
\qed
\end{Proof}
  The sign ambiguity in
  Lemma\,\ref{fact-normalized-zero-eigenfunction-1-dimension-unique}
  is fixed when one constructs $p_{\,t}(x)=p^{\,\eq}(x)\,\psi_{\,t}(x)$ with
  the properties $p^{\,\eq}(x)>0$ and
  $\psi_{\,t}(x)=\sum_{s}a^{\,(s)}(t)\phi_{\,V}^{\,(s)}(x)\geq 0$
  for all $t\in\mbbR$ and $x\in V$. 
Observe from 
Lemma\,\ref{fact-normalized-zero-eigenfunction-1-dimension-unique}
that the number of the elements of $\cN_{\,V}$ 
is calculated 
as $|\,\cN_{\,V}\,|=\# V-1$.

With these Lemmas,  the solution to \fr{detailed-balance-diffusion-equation},
$\psi_{\,t}$, is obtained as follows:
\begin{Proposition}
\label{fact-spectrum-decomposition-solution-diffusion}
(Spectrum decomposition). 
Let $\{\lambda_{\,V}^{\,(s)}\}$ be 
the totality of non-zero eigenvalues of $\Delta_{\,V}$, and
$\{\phi_{\,V}^{\,(s)}\}$ that of the corresponding eigenfunctions. Then 
\beq
\psi_{\,t}
=\sum_{s\in\cN_{\,V}}a^{\,(s)}(0)\,\e^{\ -|\,\lambda_{\,V}^{\,(s)}\,|\,t}
\phi_{\,V}^{\,(s)}+1_{\,V},
\label{decomposition-psi-eigen-basis-solution}
\eeq   
is a solution to the diffusion equations   
\fr{detailed-balance-diffusion-equation} derived from the master equations.
\end{Proposition}
\begin{Proof}
From \fr{decomposition-psi-eigen-basis}, it follows that 
$$
\inp{\frac{\dr}{\dr t}\psi_{\,t}}{\phi_{\,V}^{\,(s)}}_{\,V}
=\frac{\dr a^{\,(s)}}{\dr t},\quad\mbox{and}\quad
\inp{\Delta_{\,V}\,\psi_{\,t}}{\phi_{\,V}^{\,(s)}}_{\,V}
=\lambda_{\,V}^{\,(s)}a^{\,(s)}.
$$
With these equations and 
\fr{detailed-balance-diffusion-equation}, one has 
$$
\inp{\frac{\dr}{\dr t}\psi_{\,t}-\Delta_{\,V}\psi_{\,t}}{\phi_{\,V}^{\,(s)}}_{\,V}
=\frac{\dr a^{\,(s)}}{\dr t}-\lambda_{\,V}^{\,(s)}a^{\,(s)}
=0,
$$
from which $a^{\,(s)}(t)=a^{\,(s)}(0)\exp(\lambda_{\,V}^{\,(s)}t)$. 
Taking into account 
Remark \ref{fact-all-eigenvalues-Delta-V-positive}, one can write 
$\lambda_{\,V}^{\,(s)}=-|\lambda_{\,V}^{\,(s)}|\leq0$. Combining these arguments, 
one has
$$
\psi_{\,t}
=\sum_{s\in\cN_{\,V}}a^{\,(s)}(0)\,\e^{\ -|\,\lambda_{\,V}^{\,(s)}\,|\,t}\phi_{\,V}^{\,(s)}
+a^{\,(0)}(0)1_{\,V}.
$$
Finally it follows from 
Corollary\,\ref{fact-master-equations-detailed-balance-relaxation} 
that $a^{\,(0)}(0)$ above is unity. This completes the proof. 
\qed
\end{Proof}
Notice that 
\beq
\frac{\dr a^{\,(s)}}{\dr t}
=\lambda_{\,V}^{\,(s)} a^{\,(s)},
\quad\mbox{and}\quad
\frac{\dr}{\dr t}\psi_{\,t}
=\sum_{s}\frac{\dr a^{\,(s)}(t)}{\dr t}\phi_{\,V}^{\,(s)}
=\sum_{s}\lambda_{\,V}^{\,(s)}a^{\,(s)}(t)\phi_{\,V}^{\,(s)}
=\sum_{s\in\cN_{\,V}}\lambda_{\,V}^{\,(s)}a^{\,(s)}(t)\phi_{\,V}^{\,(s)},
\label{decomposition-psi-eigen-basis-solution-2}
\eeq
and
\beq
\psi_{\,t}
=\sum_{s\in\cN_{\,V}}a^{\,(s)}(t)\phi_{\,V}^{\,(s)}+1_{\,V}.
\label{decomposition-psi-eigen-basis-solution-3}
\eeq
An example of the spectrum decomposition of $\psi_{\,t}$ is given below.
\begin{Example}
\label{example-kinetic-ising-psi-eigen}  
(Eigen system for kinetic Ising model without spin-coupling).  
Consider Example\,\ref{example-kinetic-ising-psi}.
The matrix form of \fr{psi-kinetic-Ising} is immediately obtained as 
$$
\frac{\dr}{\dr t}
\left(
\begin{array}{c}
\psi_{\,t,\Ising}(-1)\\
\psi_{\,t,\Ising}(+1)
\end{array}
\right)
=M_{\,\Ising}
\left(
\begin{array}{c}
\psi_{\,t,\Ising}(-1)\\
\psi_{\,t,\Ising}(+1)
\end{array}
\right),\qquad \mbox{where}\qquad
M_{\,\Ising}
:=\left(
\begin{array}{cc}
-\,w_{\,-1\to+1}&w_{\,-1\to+1}\\
w_{\,+1\to -1}&-\,w_{\,+1\to-1}
\end{array}
\right).
$$
Two eigenvalues of $M_{\,\Ising}$ are obtained as 
$$
\lambda_{\,\Ising}^{\,(0)}
=0,\qquad\mbox{ and  }\qquad
\lambda_{\,\Ising}^{\,(1)}
=-\,(\,w_{\,+1\to-1}+w_{\,-1\to+1}\,).
$$
Their orthonormal eigenfunctions are 
$$
\left(
\begin{array}{c}
\phi_{\,\Ising}^{\,(0)}(-1)\\
\phi_{\,\Ising}^{\,(0)}(+1)
\end{array}
\right)
=\left(
\begin{array}{c}
1\\
1
\end{array}
\right),\qquad\mbox{and}\qquad
\left(
\begin{array}{c}
\phi_{\,\Ising}^{\,(1)}(-1)\\
\phi_{\,\Ising}^{\,(1)}(+1)
\end{array}
\right)
=c_{\,\Ising}^{\,(1)}\left(
\begin{array}{c}
w_{\,-1\to+1}\\
-\,w_{\,+1\to-1}
\end{array}
\right),
$$
where 
$$
c_{\,\Ising}^{\,(1)}
:=\left[\,(w_{\,-1\to+1})^{\,2}p_{\,\Ising}^{\,\eq}(-1)+
(w_{\,+1\to-1})^{\,2}p_{\,\Ising}^{\,\eq}(1)\right]^{-1/2}.
$$
The orthonormality is verified as 
\beqa
\inp{\phi_{\,\Ising}^{\,(0)}}{\phi_{\,\Ising}^{\,(0)}}_{\,V}
&=&\sum_{\sigma=\pm1}p_{\,\Ising}^{\,\eq}(\sigma)
=1,\non\\
\inp{\phi_{\,\Ising}^{\,(0)}}{\phi_{\,\Ising}^{\,(1)}}_{\,V}
&=&\sum_{\sigma=\pm1}\phi_{\,\Ising}^{\,(0)}(\sigma)\,
\phi_{\,\Ising}^{\,(1)}(\sigma)\,p_{\,\Ising}^{\,\eq}(\sigma)
\non\\
&=&c_{\,\Ising}^{\,(1)}\left[\,\,
w_{\,-1\to+1}\,p_{\,\Ising}^{\,\eq}(-1)
-w_{\,+1\to-1}\,p_{\,\Ising}^{\,\eq}(1)\,\right]
=0,
\non\\
\inp{\phi_{\,\Ising}^{\,(1)}}{\phi_{\,\Ising}^{\,(1)}}_{\,V}
&=&\sum_{\sigma=\pm1}\phi_{\,\Ising}^{\,(1)}(\sigma)\,
\phi_{\,\Ising}^{\,(1)}(\sigma)\,p_{\,\Ising}^{\,\eq}(\sigma)
\non\\
&=&c_{\,\Ising}^{\,(1)\,2}\left[\,(w_{\,-1\to+1})^{\,2}p_{\,\Ising}^{\,\eq}(-1)+
(w_{\,+1\to-1})^{\,2}p_{\,\Ising}^{\,\eq}(1)\right]
=1,
\non
\eeqa
where the normalization of 
$p_{\,\Ising}^{\,\eq}$ and the detailed balance condition have been used.
In Ref.\,\cite{Goto2015}, the equilibrium distribution $p_{\,\Ising}^{\,\eq}$ 
and $w_{\,\sigma\to-\sigma}$ were chosen so that they depend on parameters 
$\theta\in\mbbR$ and $\gamma\in\mbbR_{\,+}$ as 
$$
p_{\,\Ising}^{\,\eq}(\sigma;\theta)
=\frac{\exp(\theta\sigma)}{2\cosh\theta},
\qquad\mbox{and}\qquad
w_{\,\sigma\to-\sigma}(\,\theta,\gamma\,)
=\frac{\gamma}{2}
\left(1-\sigma\tanh\theta\right).
$$
The physical meaning of $\theta$ is a quantity that is 
proportional to the inverse temperature, and that of $\gamma$ is a 
characteristic relaxation time.  
The non-zero eigenvalue and its corresponding eigenfunction for this case 
are  
$$
\lambda_{\,\Ising}^{(1)}
=-\,\gamma,
\qquad\mbox{and}\qquad
\left(
\begin{array}{c}
\phi_{\,\Ising}^{(1)}(-1)\\
\phi_{\,\Ising}^{(1)}(+1)\\
\end{array}
\right)
=c_{\,\Ising}^{\,(1)}(\theta,\gamma)\,\frac{\gamma}{2}\left(
\begin{array}{c}
1+\tanh\theta\\
-1+\tanh\theta
\end{array}
\right).
$$ 
Thus, the spectrum decomposition \fr{decomposition-psi-eigen-basis-solution}
is obtained as
$$
\left(
\begin{array}{c}
  \psi_{\,\Ising,\,t}(-1)\\
  \psi_{\,\Ising,\,t}(+1)  
\end{array}
  \right)
=\check{a}_{\,\Ising}^{\,(1)}(\theta,\gamma)\,\e^{\,-\,\gamma\,t}\,
\left(
\begin{array}{c}
1+\tanh\theta\\
-1+\tanh\theta
\end{array}
\right)
+
\left(
\begin{array}{c}
1\\
1
\end{array}
\right),
$$
where
$$
\check{a}_{\,\Ising}^{\,(1)}(\theta,\gamma)
=\frac{\gamma}{2}\,c_{\,\Ising}^{\,(1)}(\theta,\gamma)\,
a_{\,\Ising}^{\,(1)}(0)
$$
with $a_{\,\Ising}^{\,(1)}(0)\geq 0$ being an initial constant for
the dynamical system 
$\dot{a}_{\,\Ising}^{\,(1)}=-\,\gamma\, a_{\,\Ising}^{\,(1)}$.  
\end{Example}

So far eigenvalues and eigenfunctions of $\Delta_{\,V}$ have been investigated. 
There exists a link between eigenvalues of $\Delta_{\,V}$ and those 
of $\Delta_{\,E}$. To state this, notations are introduced as follows.
Let $\cN_{\,E}$ be the totality of labels for non-zero eigenfunctions of 
$\Delta_{\,E}$, $\Spec^{\,\prime}\Delta_{\,V}$ and $\Spec^{\,\prime}\Delta_{\,E}$ 
the totalities of non-zero eigenvalues of $\Delta_{\,V}$ and
  that of $\Delta_{\,E}$, respectively.   
Moreover,
recall $\ker\Delta_{\,V}$, $\dim(\ker\Delta_{\,V})=1$ from  
Lemma\,\ref{fact-normalized-zero-eigenfunction-1-dimension-unique},    
and       
let 
$$
\ker\Delta_{\,V}
=\{\,\phi_{\,V}\in\Lambda^{\,0}(G)\,|\,
\Delta_{\,V}\,\phi_{\,V}=0
\,\},\qquad
\ker\Delta_{\,E}
:=\{\,\phi_{\,E}\in\Lambda^{\,1}(G)\,|\,
\Delta_{\,E}\,\phi_{\,E}
=0
\,\},
$$
$$
\ker\dr
:=\{\,\phi_{\,V}\in\Lambda^{\,0}(G)\,|\,
\dr\phi_{\,V}=0
\,\},\qquad\mbox{and}\qquad
\ker\dr^{\,\dagger}
:=\{\,\phi_{\,E}\in\Lambda^{\,1}(G)\,|\,
\dr^{\,\dagger}\phi_{\,E}=0
\,\}.
$$
    
First, one has the following:
\begin{Lemma}
\label{fact-ker-d-ker-Delta-V}
$$
\ker\dr 
=\ker\Delta_{\,V}.
$$
\end{Lemma}
\begin{Proof}
This proof can be split into two steps. First, 
$\ker\dr\subseteq\ker\Delta_{\,V}$ is shown.
Then, $\ker\Delta_{\,V}\subseteq\ker\dr$ is shown.
From these, one completes the proof.

(Proof of\quad $\ker\dr\subseteq\ker\Delta_{\,V}$ ):  
Take $\phi_{\,V}^{\,(0)}\in\ker\dr$, i.e., $\dr\phi_{\,V}^{\,(0)}=0$. 
Then it follows that 
$$
\Delta_{\,V}\,\phi_{\,V}^{\,(0)}
=-\,\dr^{\,\dagger}\dr \phi_{\,V}^{\,(0)}
=0,
$$
from which $\ker\dr\subseteq\ker\Delta_{\,V}$.

(Proof of\quad $\ker\Delta_{\,V}\subseteq\ker\dr$ ):  
Take $\phi_{\,V}^{\,(0)}\in\ker\Delta_{\,V}$, i.e., 
$\Delta_{\,V}\phi_{\,V}^{\,(0)}=0$.
Then it follows from    
$$
0=\inp{\Delta_{\,V}\,\phi_{\,V}^{\,(0)}}{\phi_{\,V}^{\,(0)}}_{\,V}
=-\inp{\dr^{\,\dagger}\dr\,\phi_{\,V}^{\,(0)}}{\phi_{\,V}^{\,(0)}}_{\,V} 
=-\inp{\dr\,\phi_{\,V}^{\,(0)}}{\dr\phi_{\,V}^{\,(0)}}_{\,E} 
\leq 0
$$
that $\dr\phi_{\,V}^{\,(0)}=0$. Thus, $\ker\Delta_{\,V}\subseteq\ker\dr$.

These two steps yield $\ker\dr=\ker\Delta_{\,V}$.
\qed
\end{Proof}
Similar to Lemma\,\ref{fact-ker-d-ker-Delta-V}, one has the following: 
\begin{Lemma}
\label{fact-ker-d-dagger-ker-Delta-E}
$$
\ker\dr^{\,\dagger} 
=\ker\Delta_{\,E}.
$$
\end{Lemma}
\begin{Proof}
A way to prove this is analogous to that of 
Lemma\,\ref{fact-ker-d-ker-Delta-V}.
\qed
\end{Proof}

Then one has the following property, 
referred to as {\it supersymmetry}\, \cite{Nakahara}:
\begin{Theorem}
\label{fact-supersymmetry}
(Supersymmetry of Laplacians).  
$$
\Spec^{\,\prime}\Delta_{\,V}
=\Spec^{\,\prime}\Delta_{\,E}.
$$
\end{Theorem}
\begin{Proof}
This proof can be split into two steps. First, 
$\Spec^{\,\prime}\Delta_{\,V}\subseteq\Spec^{\,\prime}\Delta_{\,E}$ is shown.
Then, $\Spec^{\,\prime}\Delta_{\,E}\subseteq\Spec^{\,\prime}\Delta_{\,V}$ is shown.
From these, one completes the proof.

( Proof of\quad $\Spec^{\,\prime}\Delta_{\,V}\subseteq\Spec^{\,\prime}\Delta_{\,E}$ ): 
Assume that $\Delta_{\,V}\phi_{\,V}^{\,(s)}=\lambda_{\,V}^{\,(s)}\phi_{\,V}^{\,(s)}$, 
$(s\in\cN_{\,V})$. Then, $\{\lambda_{\,V}^{\,(s)}\}=\Spec^{\,\prime}\Delta_{\,V}$.
From this assumption and Lemma\,\ref{fact-ker-d-ker-Delta-V}, 
it follows that $\dr\, \phi_{\,V}^{\,(s)}\neq 0$.
Since  
$$
\Delta_{\,E}(\,\dr\, \phi_{\,V}^{\,(s)})
=\dr\,\Delta_{\,V}\phi_{\,V}^{\,(s)}
=\lambda_{\,V}^{\,(s)}(\,\dr\, \phi_{\,V}^{\,(s)}),
\qquad\mbox{with}\quad 
\dr\, \phi_{\,V}^{\,(s)}\neq 0,  
$$
one has that $\lambda_{\,V}^{\,(s)}\in\Spec^{\,\prime}\Delta_{\,E}$ for each 
$s\in\cN_{\,V}$. 
This yields that $\Spec^{\,\prime}\Delta_{\,V}\subseteq\Spec^{\,\prime}\Delta_{\,E}$.

( Proof of\quad $\Spec^{\,\prime}\Delta_{\,E}\subseteq\Spec^{\,\prime}\Delta_{\,V}$ ): 
Assume that $\Delta_{\,E}\,\phi_{\,E}^{\,(s)}=\lambda_{\,E}^{\,(s)}\phi_{\,E}^{\,(s)}$,  
$(s\in\cN_{\,E})$. Then, $\{\lambda_{\,E}^{\,(s)}\}=\Spec^{\,\prime}\Delta_{\,E}$.
From this assumption and Lemma\,\ref{fact-ker-d-dagger-ker-Delta-E}, 
it follows that $\dr^{\,\dagger}\, \phi_{\,E}^{\,(s)}\neq 0$.
Since  
$$
\Delta_{\,V}(\,\dr^{\,\dagger}\, \phi_{\,E}^{\,(s)})
=\dr^{\,\dagger}\,\Delta_{\,E}\,\phi_{\,E}^{\,(s)}
=\lambda_{\,E}^{\,(s)}(\,\dr^{\,\dagger}\, \phi_{\,E}^{\,(s)}),
\qquad\mbox{with}\quad
\dr^{\,\dagger}\, \phi_{\,E}^{\,(s)}\neq 0, 
$$
one has that $\lambda_{\,E}^{\,(s)}\in\Spec^{\,\prime}\Delta_{\,V}$ for each 
$s\in\cN_{\,E}$. 
This yields that $\Spec^{\,\prime}\Delta_{\,E}\subseteq\Spec^{\,\prime}\Delta_{\,V}$.

These two steps yield the desired equality. 
\qed
\end{Proof}
Due to Theorem\,\ref{fact-supersymmetry}, non-zero eigenvalues 
$\lambda_{\,V}^{\,(s)}, (s\in \cN_{\,V})$ and $\lambda_{\,E}^{\,(s)},(s\in\cN_{\,E})$
are not distinguished. In what follows they are denoted $\lambda^{\,(s)}$.

\subsubsection{Expectation values}
For the case where the detailed balance conditions are not imposed, 
how expectation values are described has been argued  
in Section\,\ref{section-expectation-values-general-w}.
It is also of interest to formulate how expectation values are described
for the case that the detailed balance conditions are satisfied.
To establish such a formulation, a relation between an 
expectation value and the inner product on $\Lambda^{\,0}(G)$ is shown first.
Second, inequalities for sums involving $\psi_{\,t}$ over vertexes 
are shown.
Then some identities for such sums 
are shown.
Finally dynamical systems for expectation variables are
  constructed. 
  In general, one of the most fundamental roles in statistical mechanics
  is to link properties of macroscopic quantities
  and microscopic dynamical systems. For the case that
  macroscopic quantities can be
  identified with expectation variables, and microscopic dynamical variables
  can be identified with master equations, 
  the approaches shown below are expected to be employed in a wider class of
  systems.

The expectation value of $\cO^{\,0}\in\Lambda^{\,0}(G)$ is denoted by 
$\mbbE_{\,p_{t}}[\,\cO^{\,0}\,]$ as in \fr{expectation-value-general-p-O}.
This expectation value is written in terms of the inner product with 
\fr{detailed-balance-measures} as 
$$
\mbbE_{\,p_{t}}[\,\cO^{\,0}\,]
=\sum_{x\in V}\cO^{\,0}(x)\,p_{\,t}(x)
=\inp{\cO^{\,0}}{\psi_{\,t}}_{\,V},
$$
where $p_{\,t}(x)=p^{\,\eq}(x)\psi_{\,t}(x)$ 
in \fr{detailed-balance-split-wave-function} has been used.
The time-development of
$$
\avgg{\cO^{\,0}}{V}
:=\mbbE_{\,p_{t}}[\,\cO^{\,0}\,],
$$
for $\cO^{\,0}\in\Lambda^{\,0}(G)$ is then described by 
$$
\frac{\dr}{\dr t}\avgg{\cO^{\,0}}{V}
=\inp{\cO^{\,0}}{\dot{\psi}_{\,t}}_{\,V}
=\inp{\cO^{\,0}}{\Delta_{\,V}\psi_{\,t}}_{\,V}
=\inp{\Delta_{\,V}\,\cO^{\,0}}{\psi_{\,t}}_{\,V}.
$$
Equivalently, it follows from 
\fr{cohomology-master-equations-divergence-psi}
that
$$
\frac{\dr}{\dr t}\inp{\psi_{\,t}}{\cO^{\,0}}_{\,V}
+\inp{\ddiv\,\Pi_{\,t}}{\cO^{\,0}}_{\,V}
=0.
$$

The H function or KL-divergence
defined by 
$$
D(p_{\,t}\| p^{\,\eq})
:=\mbbE_{\,p_{t}}\left[\,\ln\,\frac{p_{\,t}}{p^{\,\eq}}\,\right]
=\sum_{x\in V}p_{\,t}(x)\ln\,\frac{p_{\,t}(x)}{p^{\,\eq}(x)}
$$
plays a central role in information theory. 
It is known that \cite{Broeck2013}
$$
\frac{\dr}{\dr t}D(p_{\,t}\| p^{\,\eq})
\,\leq\, 0.
$$
This can be proven in the present formulation as follows.
The derivative of $D(p_{\,t}\| p^{\,\eq})$ with respect to $t$ is
written in terms of  $p^{\,\eq}(x)$ and $\psi_{\,t}(x)$ as 
$$
D(p_{\,t}\| p^{\,\eq})
=\sum_{x\in V}\dot{p}_{\,t}(x)\left(1+\ln\frac{p_{\,t}(x)}{p^{\,\eq}(x)}\right)
=\sum_{x\in V}p^{\,\eq}(x)\dot{\psi}_{\,t}(x)
\left(1+\ln\psi_{\,t}(x)\right).
$$
Then applying the inequality 
$$
1+\ln \xi\leq \xi,\qquad \xi\geq0
$$
and \fr{detailed-balance-diffusion-equation}, one has
$$
\frac{\dr}{\dr t}D(p_{\,t}\| p^{\,\eq})
\leq\sum_{x\in V}p^{\,\eq}(x)\dot{\psi}_{\,t}(x)
\psi_{\,t}(x)
=\inp{\dot{\psi}_{\,t}}{\psi_{\,t}}_{\,V}
=\inp{\Delta_{\,V}\psi_{\,t}}{\psi_{\,t}}_{\,V}.
$$
Finally from
$\inp{\dr\psi_{\,t}}{\dr\psi_{\,t}}_{V}\geq0$,  
one has the desired inequality:
$$
\frac{\dr}{\dr t}D(p_{\,t}\| p^{\,\eq})
\leq
-\inp{\dr\psi_{\,t}}{\dr\psi_{\,t}}_{\,V}
\leq0.
$$
It is then of interest to explore similar inequalities for sums 
involving $\psi_{\,t}$ over vertexes. 
Define $S_{\,t}\in\Lambda^{\,0}(G)$ to be  
$$
S_{\,t}
:=-\ln \psi_{\,t}.
$$
A significance of $S_{\,t}$ is discussed after showing the
following Proposition:
\begin{Proposition}
\label{fact-inequality-S}
  (Inequality for $S_{\,t}$).  
$$
\inp{S_{\,t}}{1_{\,V}}_{\,V}
\geq 0,\qquad \forall\,t\in\mbbR.
$$
\end{Proposition}
\begin{Proof}
By applying the inequality
$$
\exp(-\,c)
\geq 1-c,\qquad c\in\mbbR, 
$$
to the equality 
\beq
1=\sum_{x\in V}p_t(x)
=\sum_{x\in V}p^{\,\eq}(x)\exp(-S_{\,t}(x))
=\inp{\exp(-S_{\,t})}{1_{\,V}}_{\,V},
\label{integral-fluctuation-theorem-S}
\eeq
one has 
$$
\inp{S_{\,t}}{1_{\,V}}_{\,V}
\geq 0.
$$
\qed
\end{Proof}
  From \fr{integral-fluctuation-theorem-S},
  the quantity $S_{\,t}$ obeys an integral fluctuation theorem
  in the sense that\,\cite{Esposito2010},    
  $$ 
  \mbbE_{\,p^{\,\eq}}[\,\e^{\,-S_{\,t}}\,]
  =1.
  $$
Fluctuation theorems hold in far from equilibrium states, and thus  
they are used as basic tools in the study of nonequilibrium
statistical mechanics\,\cite{Seifert2008}.
Note that Proposition\,\ref{fact-inequality-S} can be written as
$\mbbE_{\,p^{\,\eq}}[\,S_{\,t}\,]\geq 0$. 
Related to this, it is shown below how the time-dependence of
$\mbbE_{\,p_{t}}[\,S_{\,t}\,]$ is related to eigenvalues of the derived
diffusion equations.
To this end, define
\beq
H[\psi_{\,t}]
:=\mbbE_{\,p_{t}}[\,S_{\,t}\,]
=\mbbE_{\,p_{t}}[\,-\ln\,\psi_{\,t}\,]
=\inp{S_{\,t}}{\psi_{\,t}}_{\,V}.
\label{entropy-psi}
\eeq
Then one has the following.
\begin{Proposition}
\label{fact-inequality-H-psi}
  (Inequality for time-derivative of $H[\psi_{\,t}]$).  
$$
\frac{\dr}{\dr t}H[\psi_{\,t}]
\geq \sum_{s}|\,\lambda^{\,(s)}\,|\,(\,a^{\,(s)}(t)\,)^{2}
>0.
$$
\end{Proposition}
\begin{Proof}
Substituting the inequality
$$
1+\ln\psi_{\,t}(x)
\leq \psi_{\,t}(x),\qquad\forall x\in V
$$
and \fr{detailed-balance-diffusion-equation} into 
\fr{entropy-psi},  one has  
\beqa
-\,\frac{\dr}{\dr t}H[\psi_{\,t}]
&=&\sum_{x\in V}
p^{\,\eq}(x)\frac{\dr}{\dr t}\left[\,
\psi_{\,t}(x)\ln\psi_{\,t}(x)
\,\right]
=\sum_{x\in V}p^{\,\eq}(x)\left(1+\ln\psi_{\,t}(x)\right)\dot{\psi}_{\,t}(x)
\non\\
&\leq& 
\sum_{x\in V}p^{\,\eq}(x)\,\psi_{\,t}(x)\,\dot{\psi_{\,t}}(x)
=\inp{\psi_{\,t}}{\Delta_{\,V}\psi_{\,t}}_{\,V}.
\non
\eeqa
From this and \fr{decomposition-psi-eigen-basis}, it follows that 
$$
-\,\frac{\dr}{\dr t}H[\psi_{\,t}]
\leq 
\sum_{s}\sum_{s^{\,\prime}}\lambda^{\,(s^{\,\prime})}\,a^{\,(s)}(t)a^{\,(s^{\,\prime})}(t)\,
\inp{\phi_{\,V}^{\,(s)}}{\phi_{\,V}^{\,(s^{\,\prime})}}_{\,V}
=\sum_{s}\lambda^{\,(s)}\,(\,a^{\,(s)}(t)\,)^{\,2},
$$ 
from which 
$$
\frac{\dr}{\dr t}H[\psi_{\,t}]
\geq
-\sum_{s}\lambda^{\,(s)}\,(\,a^{\,(s)}(t)\,)^{\,2}
=-\sum_{s\in\cN_{\,V}}\lambda^{\,(s)}\,(\,a^{\,(s)}(t)\,)^{\,2}.
$$
Since $\lambda^{(s)}=-|\lambda^{(s)}|<0$ for $s\in\cN_{\,V}$ 
(See Remark \ref{fact-all-eigenvalues-Delta-V-positive}), 
one has that 
$$
\frac{\dr}{\dr t}H[\psi_{\,t}]
\geq
\sum_{s\in\cN_{\,V}}|\,\lambda^{\,(s)}\,|\,(\,a^{\,(s)}(t)\,)^{\,2}
>0.
$$
\qed
\end{Proof}
To construct a dynamical system for 
an expectation variable $\avgg{\cO^{\,0}}{V}=\mbbE_{\,p_{t}}[\,\cO^{\,0}\,]$ with 
the spectrum decomposition,
one starts with 
\beq
\frac{\dr}{\dr t}\avgg{\cO^{\,0}}{V} 
  =\inp{\cO^{\,0}}{\dot{\psi}}_{\,V}
=\sum_{s\in\cN_{\,V}}\frac{\dr a^{\,(s)}(t)}{\dr t}\sum_{x\in V}
  \cO^{\,0}(x)\,\phi_{\,V}^{\,(s)}(x)\,p^{\,\eq}(x),\qquad 
\mbox{with}\quad \dot{a}^{\,s}=\lambda^{\,(s)}a^{\,s},
\label{expectation-value-equation-spectrum-decomposition-derivative}
\eeq
where \fr{decomposition-psi-eigen-basis-solution-3} has been used.
The right hand side of the equation above
 might not be written in terms of a function of
\beq
\avgg{\cO^{\,0}}{V}
=\sum_{x\in V}\cO^{\,0}(x)\left[\sum_{s\in\cN_{\,V}}a^{\,(s)}(t)
  \,\phi_{\,V}^{\,(s)}(x)+1\right]\,p^{\,\eq}(x).
\label{expectation-value-equation-spectrum-decomposition}
\eeq
For this reason, it is not 
obvious whether there exists a closed 
dynamical system for $\avgg{\cO^{\,0}}{V}$. 
Nevertheless,
for some simple systems, such closed dynamical systems can be constructed from
master equations. For example, one has the following:
\begin{Proposition}
\label{fact-system-from-master-equations-with-a-unique-non-trivial-eigenvalue}
  (Dynamical system for master equations with a unique non-trivial eigenvalue).  
Consider the case where $|\,\cN_{\,V}\,|=1$. In this case, one has 
the dynamical system on $\mbbR$ 
\beq
\frac{\dr}{\dr t}\avgg{\cO^{\,0}}{V}
=\lambda^{\,(1)}\,\left(\,\avgg{\cO^{\,0}}{V}-\avgg{\cO^{\,0}}{V}^{\,\eq}
\,\right),\quad\mbox{where}\quad
\avgg{\cO^{\,0}}{V}^{\,\eq}
:=\sum_{x\in V}\cO^{\,0}(x)\,p^{\,\eq}(x).
\label{fact-dynamical-systems-non-trivial-eigenvalue-1}
\eeq
\end{Proposition}
\begin{Proof}
Since $|\,\cN_{\,V}\,|=1$,
one has from \fr{expectation-value-equation-spectrum-decomposition-derivative}  
that
$$
\frac{\dr}{\dr t}\avgg{\cO^{\,0}}{V} 
=\lambda^{\,(1)}\,a^{\,(1)}(t)\sum_{x\in V}
  \cO^{\,0}(x)\,\phi_{\,V}^{\,(1)}(x)\,p^{\,\eq}(x),
  $$
and from \fr{expectation-value-equation-spectrum-decomposition} that
$$
\avgg{\cO^{\,0}}{V}
=\sum_{x\in V}\cO^{\,0}(x)\left[\,a^{\,(1)}(t)
  \,\phi_{\,V}^{\,(1)}(x)+1\,\right]\,p^{\,\eq}(x)
=a^{\,(1)}(t)\sum_{x\in V}\cO^{\,0}(x)
  \,\phi_{\,V}^{\,(1)}(x)p^{\,\eq}(x)+\avgg{\cO^{\,0}}{V}^{\,\eq}.
$$
Substituting the second equation into the first one, one completes the proof. 
\qed
\end{Proof}

The solution to \fr{fact-dynamical-systems-non-trivial-eigenvalue-1}
is obtained as  
$$
\avgg{\cO^{\,0}}{V}(t)
=\avgg{\cO^{\,0}}{V}^{\,\eq}+\left(\,
\avgg{\cO^{\,0}}{V}(0)-\avgg{\cO^{\,0}}{V}^{\,\eq}\,\right)\,\e^{\,\lambda^{\,(1)}\,t}.
$$
It follows from this expression with $\lambda^{\,(1)}< 0$ that the 
equilibrium state for $\avgg{\cO^{\,0}}{V}$ is indeed uniquely realized as $t\to\infty$: 
$$
\lim_{t\to\infty}\avgg{\cO^{\,0}}{V}(t)
=\avgg{\cO^{\,0}}{V}^{\,\eq}.
$$  
The following is an example associated with Proposition\,\ref{fact-system-from-master-equations-with-a-unique-non-trivial-eigenvalue}:
\begin{Example}
(Dynamical system for magnetization driven by kinetic Ising model).
Consider Examples\,\ref{example-kinetic-ising-psi} and
\ref{example-kinetic-ising-psi-eigen}.  In this analysis
$p_{\,\Ising}^{\,\eq}(\sigma;\theta)$ is abbreviated as $p_{\,\Ising}^{\,\eq}(\sigma)$.   
Choose $\cO^{\,0}\in\Lambda^{\,0}(G)$ to be 
$$
\cO^{\,0}(\sigma)
=\sigma.
$$
The expectation variable of this $\cO^{\,0}$ is time-dependent magnetization,
expressed as 
\beqa
\avgg{\sigma}{V}
&=&\sum_{\sigma=\pm1}\sigma\,
p_{\,\Ising}^{\,\eq}(\sigma)\,
\left(\,
  a_{\,\Ising}^{\,(1)}(0)\,\e^{\,-\,\gamma\,t}\,\phi_{\,\Ising}^{\,(1)}(\sigma)
+\phi_{\,\Ising}^{\,(0)}(\sigma)
  \, \right)
\non\\
&=&
a_{\,\Ising}^{\,(1)}(0)\,\e^{\,-\,\gamma\,t}\,\left(
  p_{\,\Ising}^{\,\eq}(+1)\,\phi_{\,\Ising}^{\,(1)}(+1)
  -p_{\,\Ising}^{\,\eq}(-1) \,\phi_{\,\Ising}^{\,(1)}(-1)
  \right)
+\left(\,p_{\,\Ising}^{\,\eq}(+1)-p_{\,\Ising}^{\,\eq}(-1)\,\right).
\non
\eeqa
Its derivative with respect to $t$ is calculated, and that derivative 
can be
expressed in terms of $\avgg{\sigma}{V}$ as 
\beqa
\frac{\dr}{\dr t}\avgg{\sigma}{V}
&=&-\,\gamma\,a_{\,\Ising}^{\,(1)}(0)\,\e^{\,-\,\gamma\,t}\,\left(\,
  p_{\,\Ising}^{\,\eq}(+1)\,\phi_{\,\Ising}^{\,(1)}(+1)
  -p_{\,\Ising}^{\,\eq}(-1) \,\phi_{\,\Ising}^{\,(1)}(-1)
 \, \right)
\non\\
&=&-\,\gamma\,\left[\,\avgg{\sigma}{V}
  -\left(\,
p_{\,\Ising}^{\,\eq}(+1)-p_{\,\Ising}^{\,\eq}(-1)
\,  \right)\,\right].
\non
\eeqa
Substituting $p_{\,\Ising}^{\,\eq}(\sigma)=\e^{\,\theta\,\sigma}/(2\cosh\theta)$
into the equation above, one has the closed dynamical system for
$\avgg{\sigma}{V}$ 
$$
\frac{\dr}{\dr t}\avgg{\sigma}{V}
=-\,\gamma\,\left(\,\avgg{\sigma}{V}-\tanh\theta\,\right),
$$
which is a dynamical system studied in Ref.\,\cite{Goto2015}.
\end{Example}    

There exist closed dynamical systems 
for expectation variables $\mbbE_{\,p_{t}}[\,\cO^{\,0}\,]$ with some 
appropriately chosen $\cO^{\,0}$. 
Such systems are given as follows:
\begin{Proposition}
\label{fact-dynamical-systems-1}
(Dynamical systems for expectation variables 1). 
Consider the case where 
$$
\cO^{\,0}
=\phi_{\,V}^{\,(s)},\qquad s\in\cN_{\,V}. 
$$
Then, 
$$
\avgg{\phi_{\,V}^{\,(s)}}{V}(t)
:=\mbbE_{\,p_{t}}[\,\phi_{\,V}^{\,(s)}\,],
$$
follows the dynamical system 
\beq
\frac{\dr}{\dr t}\avgg{\phi_{\,V}^{\,(s)}}{V}
=\lambda^{\,(s)}\avgg{\phi_{\,V}^{\,(s)}}{V}.
\label{closed-dynamical-system-1}
\eeq
\end{Proposition}
\begin{Proof}
It follows from \fr{decomposition-psi-eigen-basis-solution-2} and
\fr{decomposition-psi-eigen-basis} that 
$$
\avgg{\phi_{\,V}^{\,(s)}}{V}
=\inp{\phi_{\,V}^{\,(s)}}{\psi_{\,t}}_{\,V}
=a^{\,(s)}(t).
$$
From this and \fr{decomposition-psi-eigen-basis-solution-2}, 
one has the closed form 
$$
\frac{\dr}{\dr t}\avgg{\phi_{\,V}^{\,(s)}}{V}
=\dot{a}^{\,(s)}(t)
=\lambda^{\,(s)}\avgg{\phi_{\,V}^{\,(s)}}{V}.
$$
\qed
\end{Proof}
  To show how \fr{closed-dynamical-system-1} is linked to another dynamical
  system in another scientific literature,
it is shown below that the diffusion equations derived from 
master equations \fr{detailed-balance-diffusion-equation} 
are used to facilitate the derivation of a dynamical system discussed
in information geometry. 
Here information geometry is a geometrization of mathematical 
statistics\,\cite{Amari-Nagaoka}. 

In the context of dynamical systems theory in information geometry, 
the dynamical system 
\beq
\frac{\dr}{\dr t}\eta_{\,j}
=-\,(\eta_{\,j}-\eta_{\,j}^{\,\prime}\,),\qquad j=1,\ldots,d
\label{fujiwara-amari-system}
\eeq
has been studied in  
\cite{Nakamura1994,Fujiwara1995,Noda2016,Goto2016}.
The variables $\{\eta_{\,j}\}$ in 
\fr{fujiwara-amari-system}  correspond to 
the expectation values for 
an exponential family of probability distribution functions,
  $\{\eta_{\,j}^{\,\prime}\}$ are constants, and $d$ is the number of canonical parameters of the exponential family.
  In Ref.\,\cite{Fujiwara1995},  
\fr{fujiwara-amari-system} was constructed as a gradient flow
by introducing the so-called divergence function.
By contrast, 
how to establish 
\fr{fujiwara-amari-system} in terms of probability distribution functions 
obeying master equations has remained unclear.

Introduce a dynamical system that is slightly different from
\fr{fujiwara-amari-system}
\beq
\frac{\dr\eta_{\,j}}{\dr t^{\,(j)}}
=-\,(\eta_{\,j}-\eta_{\,j}^{\,\prime}\,),\qquad j=1,\ldots,d
\label{fujiwara-amari-system-modified}
\eeq
with some scaled time variables $t^{\,(j)}$.  
It is shown below how \fr{closed-dynamical-system-1}
is linked to \fr{fujiwara-amari-system-modified}.
\begin{Proposition}
The dynamical system \fr{closed-dynamical-system-1}
yields \fr{fujiwara-amari-system-modified}.
\end{Proposition}
\begin{Proof}
  First, introduce $\wt{\eta}_{\,s}$ and $t^{\,(s)}$ so that  
$$
\avgg{\phi_{\,V}^{\,(s)}}{V}(t)
=\wt{\eta}_{\,s}(\,t^{\,(s)}\,(t)\,),\qquad
t^{\,(s)}(t)
:=|\,\lambda^{\,(s)}\,|\,t,
$$
from which 
$$
\frac{\dr}{\dr t}\avgg{\phi_{\,V}^{\,(s)}}{V}(t)
=|\,\lambda^{\,(s)}\,|\,\frac{\dr \wt{\eta}_{\,s}}{\dr t^{\,(s)}}.
$$
Substituting this into  \fr{closed-dynamical-system-1}, one has
$$
\frac{\dr \wt{\eta}_{\,s}}{\dr t^{\,(s)}}
=-\,\wt{\eta}_{\,s}.
$$
Second, introduce $\eta_{\,s}(\,t^{\,(s)}\,)$ and constant
$\eta_{\,s}^{\,\prime}$ so that
$$
\wt{\eta}_{\,s}\left(t^{\,(s)}\right)
=\eta_{\,s}\left(t^{\,(s)}\right)-\eta_{\,s}^{\,\prime}.
$$
The variable $\eta_{\,s}$ obeys 
$$
\frac{\dr \eta_{\,s}}{\dr t^{\,(s)}}
=-\,\left(\eta_{\,s}-\eta_{\,s}^{\,\prime}\right).
$$
This is \fr{fujiwara-amari-system-modified} with $s$ being replaced with $j$.
\qed  
\end{Proof}

Similar to Proposition\,\ref{fact-dynamical-systems-1},
closed equations on $\Lambda^{\,1}(G)$ 
are obtained  as follows.
Define  for $\cO^{\,1}\in\Lambda^{\,1}(G)$, 
$$
\avgg{\cO^{\,1}}{E}
:=\inp{\cO^{\,1}}{\dr\psi_{\,t}}_{\,E}.
$$ 
Differentiating this equation with respect to $t$ and using  
\fr{detailed-balance-split-wave-function-edge}, one has 
$$
\frac{\dr}{\dr t}\avgg{\cO^{\,1}}{E}
=\inp{\cO^{\,1}}{\Delta_{\,E}\,\dr\psi_{\,t}}_{\,E}
=\inp{\Delta_{\,E}\,\cO^{\,1}}{\dr\psi_{\,t}}_{\,E}.
$$
Then one has the following:
\begin{Proposition}
\label{fact-dynamical-systems-2}
(Dynamical systems for expectation variables 2). 
Consider the case where 
$$
\cO^{\,1}=\dr \phi_{\,V}^{\,(s)},\qquad s\in\cN_{\,E}.
$$
Then, one has 
the closed dynamical system
$$
\frac{\dr}{\dr t}\avgg{\dr\phi_{\,V}^{\,(s)}}{E}
=\lambda^{\,(s)}\avgg{\dr\phi_{\,V}^{\,(s)}}{E}.
$$
\end{Proposition}
\begin{Proof}
It follows from 
that 
$$
\frac{\dr}{\dr t}\avgg{\dr\phi_{\,V}^{\,(s)}}{E}
=\inp{\Delta_{\,E}\,\dr\phi_{\,V}^{\,(s)}}{\dr\psi_{\,t}}_{\,E}
=\inp{\lambda^{\,(s)}\dr\phi_{\,V}^{\,(s)}}{\dr\psi_{\,t}}_{\,E}
=\lambda^{\,(s)}\avgg{\dr\phi_{\,V}^{\,(s)}}{E}.
$$
\qed
\end{Proof}
\section{Conclusions}
\label{sec-conclusions}
This paper offers a viewpoint that master equations employed in 
nonequilibrium statistical mechanics can be analyzed in discrete geometry.
As the main theorem in this paper,
when the detailed balance conditions are satisfied, 
master equations induce diffusion equations, where Laplacians are associated
with chosen  inner products. These diffusion equations
can analytically be solved and then explicit solutions have been obtained. 
With the self-adjoint property of the Laplacians,
the orthonormal decomposition of the solution and closed equations for
expectation variables have been obtained. 
Furthermore, the coboundary operator and its adjoint enable us to simplify
various calculations including expectation values, and to show the so-called
supersymmetry for two classes of Laplacians.
Moreover, to illustrate how to apply these, some examples have been shown. 

The significance of this discrete geometric description
of master equations is argued from a mathematical viewpoint as follows.   
Although various graph theoretic approaches
to master equations exist in the literature,
none of them have yielded reduced equations such as diffusion equations
without any approximation. A crucial point of the derivation of
the diffusion equations is to introduce 
freedom to choose weights defining inner products for functions on graphs. 
Here the introduction of weights, or equivalently the additional freedom,
has been incorporated in discrete geometry\,\cite{Sunada2013}, and  
has then induced wider classes of Laplacians on graphs 
whose eigenvalues are real due to the self-adjoint property. The usefulness of
this self-adjoint property of Laplacians has been employed throughout 
this paper.

  The significance of the present study is argued from a physical
  viewpoint as follows.
  Given master equations satisfying the detailed balance conditions,
  diffusion equations have been obtained by introducing new variables. 
  Since diffusion equations are ubiquitous in physics,
  insight can be gained from the physical literature.
  For instance, such insight is to
  identify the time-development of master equations
  with how an ink is spread in a fluid.   
  It is one of relaxation processes in nonequilibrium systems,
  and this time-development of master equations
  can be expressed as the sum of exponential decays with real
  decay rates. Thus, oscillatory behavior in time
  is not observed in the new variables. 
  Meanwhile, inner products have been introduced by choosing measures
  in this paper. Choosing these measures has provided
  preferred or biased ways to
  describe the time-development of the master equations. Hence, other choices
  of measures could be suitable for master equations with other conditions.  
  In addition, the use of inner products is similar to that 
  in quantum mechanics\,\cite{Baez2018}. This similarity is expected to be
  useful for introducing notions developed in quantum mechanics to the study of
  master equations.  
  It should be mentioned that
one of the most fundamental roles in statistical mechanics
  is to link properties of macroscopic quantities
  and microscopic dynamical systems. When 
  macroscopic quantities can be
  identified with expectation variables, and microscopic dynamical variables
  can be identified with master equations, 
  the approaches to construct dynamical systems for expectation variables
  shown in this paper 
  are expected to be used in a wider class of systems.

There are numbers of extensions that follow from this study.
One of them is to apply the present approach to quantum systems, 
chemical reactions, and economical systems.
Another example is to consider the case where the
detailed balance conditions are not satisfied\,\cite{Sakai2013}. 
Moreover,
some infinite graph case and discrete-time case should be addressed, and
their applications to mathematical engineering, including
Monte-Carlo simulations, should be considered as future works.

\subsection*{Acknowledgments}
S.G. was partially supported by JSPS (KAKENHI) 
Grant No. JP19K03635 and is 
also grateful to Shuhei MANO for fruitful discussions.
H.H. was partially supported by JSPS (KAKENHI)
Grant No. JP17H01793. 
\newpage


\end{document}